\documentclass[english, 10pt, letterpaper]{article} 
\usepackage{tcolorbox} 
\usepackage{xcolor} 
\definecolor{myred}{RGB}{153, 0, 0} 
\definecolor{mygreen}{RGB}{0, 70, 0} 
\definecolor{myblue}{RGB}{0, 102, 204} 
\usepackage{cite} 
\usepackage[T1]{fontenc} 
\usepackage[left = 0.59in, right = 0.59in, top = 0.89in, bottom = 0.72in]{geometry} 
\usepackage{wrapfig} 
\usepackage{graphicx} 
\usepackage{authblk} 
\usepackage[hidelinks]{hyperref} 
\usepackage{orcidlink} 
\usepackage{balance} 
\usepackage[misc]{ifsym} 
\usepackage{amssymb} 
\usepackage{amsmath} 
\usepackage{gensymb} 
\usepackage{babel} 
\usepackage{indentfirst} 
\usepackage{setspace} 
\usepackage{lettrine} 
\usepackage{fancyhdr} 
\usepackage{stackengine} 
\usepackage{caption} 
\providecommand{\indexterms}[1]{\small{\textbf{\textit{Index Terms---}}}#1} 
\providecommand{\abst}[1]{\textbf{\textit{Abstract---}}#1} 
\usepackage[scaled = 1]{helvet} 
\renewcommand\cite[1]{{\color{mygreen}[\citenum{#1}]}} 
\usepackage{microtype} 
\title{\Huge\textbf{Frequency Synchronization Circuit Model for Analog Vector-Matrix Multiplication in the Frequency Domain}}
\author[1, $*$]{\textbf{Taeju~Lee}{\,}\orcidlink{0000-0001-6304-3073}}
\author[2, $\dagger$]{\textbf{Scott~T.~Habermehl}{\,}\orcidlink{0009-0006-2970-4580}}
\author[3, $\ddagger$]{\textbf{Timothy~W.~Caplice}{\,}\orcidlink{0000-0003-3561-8436}}
\author[2, 4, 5, $\S$]{\textbf{Michael~L.~Roukes}{\,}\orcidlink{0000-0002-2916-6026}}
\author[3, 6, $\P$]{\textbf{Philip~X.-L.~Feng}{\,}\orcidlink{0000-0002-1083-2391}}
\author[1, 7, 8, $\|$]{\textbf{Kenneth~L.~Shepard}{\,}\orcidlink{0000-0003-0665-6775}}

\affil[1]{\small{Department of Electrical Engineering, Columbia University, New York, NY 10027, USA}}
\affil[2]{\small{Department of Physics, California Institute of Technology, Pasadena, CA 91125, USA}}
\affil[3]{\small{Department of Physics, University of Florida, Gainesville, FL 32611, USA}}
\affil[4]{\small{Department of Applied Physics, California Institute of Technology, Pasadena, CA 91125, USA}}
\affil[5]{\small{Department of Bioengineering, California Institute of Technology, Pasadena, CA 91125, USA}}
\affil[6]{\small{Department of Electrical and Computer Engineering, University of Florida, Gainesville, FL 32611, USA}}
\affil[7]{\small{Department of Biomedical Engineering, Columbia University, New York, NY 10027, USA}}
\affil[8]{\small{Department of Neurological Surgery, Columbia University, New York, NY 10032, USA}\vspace{10pt}}
\affil[$*$]{Correspondence to{\,}{\Letter}{\,}taeju.leo.lee@gmail.com}
\affil[$\dagger$]{shaberme@caltech.edu}
\affil[$\ddagger$]{t.caplice@ufl.edu}
\affil[$\S$]{roukes@caltech.edu}
\affil[$\P$]{philip.feng@ufl.edu}
\affil[$\|$]{shepard@ee.columbia.edu}
\date{} 

\begin{document}
\onecolumn 
\maketitle
\pagestyle{fancy}
\fancyhf{}
\lhead{\footnotesize{LEE \textit{et al.}: FREQUENCY SYNCHRONIZATION CIRCUIT MODEL FOR ANALOG VMM IN THE FREQUENCY DOMAIN}}
\cfoot{\thepage} 

\begin{center}
\begin{tcolorbox}[colframe = black!0!white, colback = black!0!white, coltext = black, height = 7cm, width = 14cm, after = \vspace{0pt}]
\noindent\small{\abst{\textbf{This work introduces a frequency synchronization circuit model for analog vector-matrix multiplication in the frequency domain. The frequency synchronization is implemented by coupling oscillators through all-to-all connections. Each oscillator is implemented using complementary cross-coupled transconductance pairs and RC high-pass filters. In this work, the frequency synchronization is achieved not only using oscillators alone but also through a configuration in which oscillators and resonator models (i.e., MEMS/NEMS resonators) are combined. While addressing the frequency synchronization of all the coupled oscillators, a partial synchronization is also discussed by controlling the coupling strength of a selected subset of oscillators. This frequency synchronization analysis focuses on the small-signal analysis using analog circuit and resonator models, thereby enabling an understanding of the synchronized frequency defined by the parameters of the analog circuit and resonator models.}}}

\singlespacing\noindent\small{\indexterms{\textbf{Analog computing, analog MAC, coupled oscillators, frequency synchronization, frequency domain computing, multiply-accumulate operation, oscillatory neural network, vector-matrix multiplication.}}}
\end{tcolorbox}
\vfill{ 
\begin{center}
\begin{tcolorbox}[colframe = black!0!white, colback = black!0!white, coltext = black, height = 1cm, width = 18cm]
\noindent\rule{\textwidth}{1pt} 
\raggedleft\footnotesize{\textit{Technical Reports}{\,}\copyright{\,}2026 All Authors. All Rights Reserved.} 
\end{tcolorbox}
\end{center}
} 
\end{center}

\rfoot{\footnotesize{\textit{Technical Reports}}} 
\newpage
\tableofcontents

\newpage
\twocolumn
\section{\color{myblue}\Large{I}\large{NTRODUCTION}}
\lettrine[findent = 0pt, nindent = 0pt]{\textbf{E}}{\textbf{ncoding}} information into the frequency domain is widely employed in sensor interfaces \cite{Wang2013sensor}\text{\color{mygreen}--}\cite{Jiang2017sensor} and wireless communications \cite{Chang1966RF}\text{\color{mygreen}--}\cite{Myung2006RF}. In sensor interfaces, the dynamic range of a front-end channel can be extended by quantizing an analog input into frequency or phase variations \cite{Jiang2017sensor}, \cite{Cardes2018sensor}\text{\color{mygreen}--}\cite{Pochet2021sensor}. In reactance sensor-based applications, a sensor input can be directly quantized into the frequency of an oscillator \cite{Wang2013sensor}, \cite{Chien2016sensor}. Frequency modulation is required not only for wireless communication \cite{Chang1966RF}\text{\color{mygreen}--}\cite{Myung2006RF} but also for the energy-efficient operation of computing systems \cite{Luo2023DFS}\text{\color{mygreen}--}\cite{Ge2013DFS}. Therefore, frequency-based data processing and system operation play a crucial role in the fields of scientific instrumentation and high-performance computing.

\subsection{\color{myblue}Prior Works for Computing Engines}
The Go match between the program AlphaGo and player Lee Sedol sparked a sensational wave of interest in an artificial intelligence (AI) system \cite{Silver2016AI}, \cite{Silver2017AI}. Then, the emergence of the AI architecture \cite{Vaswani2017AI}, Transformer, has become the foundation of modern AI services, including ChatGPT \cite{Radford2018AI}\text{\color{mygreen}--}\cite{OpenAI2023AI}, Claude \cite{Bai2022AI}, Gemini \cite{Gemini2023AI}, Copilot \cite{Chen2021AI}, etc. To support these AI services, high-performance computing engines such as NVIDIA GPUs \cite{NVIDIA2022GPU}\text{\color{mygreen}--}\cite{NVIDIA2025GPU} and Google TPUs \cite{Jouppi2017TPU}\text{\color{mygreen}--}\cite{Jouppi2023TPU} have been deployed. In these hardware systems, vector-matrix multiplication (VMM) is one of the key functions for executing deep neural networks (DNNs) serving as the foundation of AI. \footnote{Vector-matrix multiplication is implemented through multiple multiply-accumulate (MAC) operations.} As computing workloads required for advanced AI models increase, the logic gate-based VMM experiences increasing latency and energy costs between processing and memory units, known as the von Neumann bottleneck. To overcome this phenomenon, VMM can be implemented through analog in-memory computing (AIMC) \cite{Burr2021AIMC}, which is achieved by combining Ohm's Law and Kirchhoff's Current Law based on non-volatile memory (NVM) units such as phase-change memory (PCM) \cite{Burr2010PCM}\text{\color{mygreen}--}\cite{Ambrogio2023PCM} and resistive random-access memory (ReRAM/RRAM) \cite{Ielmini2018RRAM}\text{\color{mygreen}--}\cite{Wan2022RRAM}.

\subsection{\color{myblue}Proposed Concept}
\begin{figure}[h!]
\centering\includegraphics[scale = 0.55]{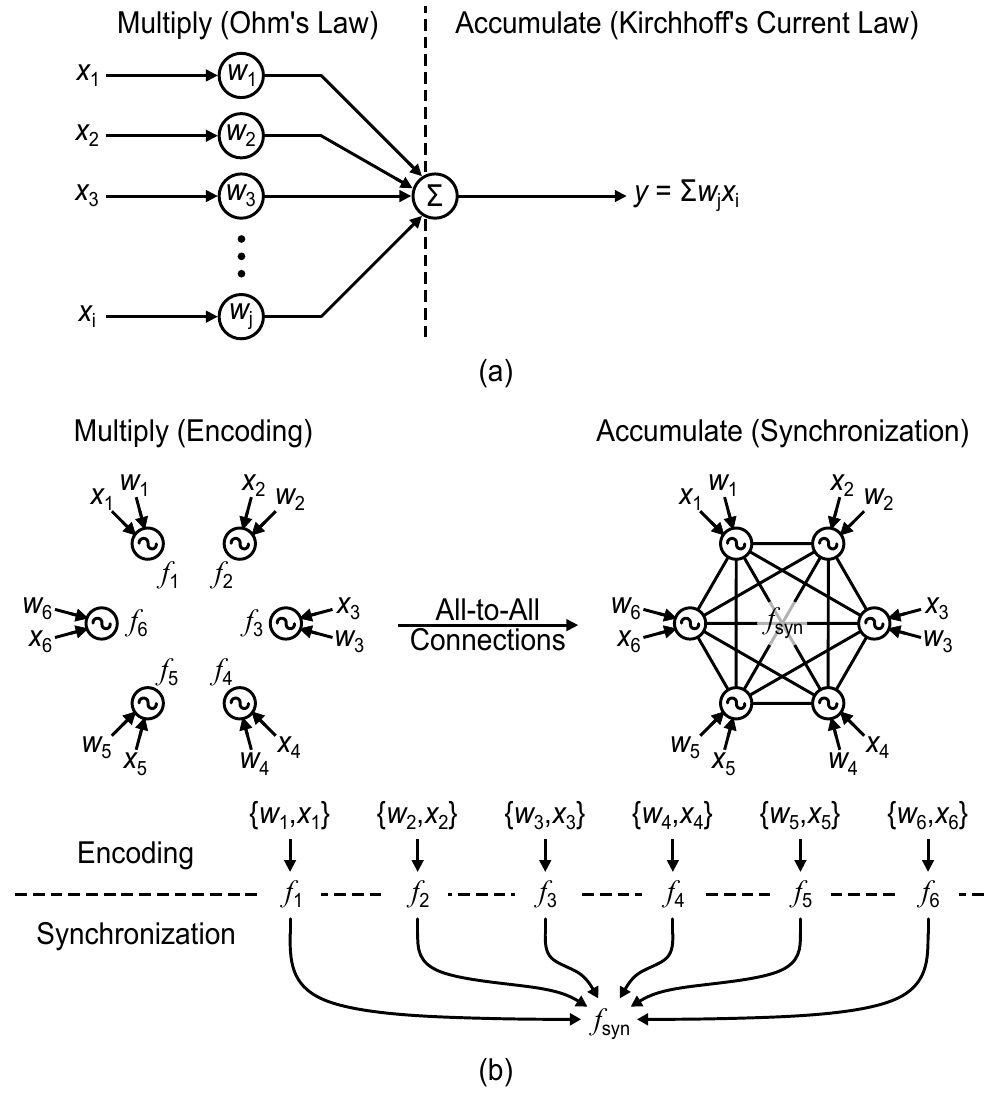}
\captionof{figure}{{\color{myred}\textbf{Analog VMM.}} (a) Conventional approach using Ohm's Law and Kirchhoff's Current Law. (b) Proposed approach using frequency encoding and synchronization.}\label{fig. 1}
\end{figure}

In contrast to AIMCs based on Ohm's Law and Kirchhoff's Current Law for VMM and AI computing as shown in Fig. \ref{fig. 1}(a) \cite{Burr2021AIMC}\text{\color{mygreen}--}\cite{Wan2022RRAM}, this work proposes a frequency synchronization circuit model for performing analog VMM in the frequency domain, as shown in Fig. \ref{fig. 1}(b). Given an input $x_{i}$ and a weight $w_{j}$, Fig. \ref{fig. 1}(a) shows the conventional analog VMM that employs Ohm's Law for multiplication, $w_{j}x_{i}$, and Kirchhoff's Current Law for accumulation, ${\Sigma}w_{j}x_{i}$.

To conduct analog VMM in the frequency domain, $x_{i}$ and $w_{j}$ are encoded into a frequency by an oscillation unit, corresponding to multiplication based on Ohm's Law. Then, the encoded frequencies are synchronized through all-to-all connections among all units, corresponding to accumulation based on Kirchhoff's Current Law.

Fig. \ref{fig. 1}(b) illustrates the operation of multiplication (encoding) and accumulation (synchronization) using 6 oscillation units. In the encoding phase, the units take datasets $\{w_{1},x_{1}\}$ through $\{w_{6},x_{6}\}$ as inputs and return the frequencies $f_{1}$ through $f_{6}$, e.g., $\{w_{1},x_{1}\}{\rightarrow}f_{1}$ and $\{w_{6},x_{6}\}{\rightarrow}f_{6}$. In the synchronization phase, the units are connected in an all-to-all manner so that all the individual frequencies are updated to a synchronized frequency $f_{syn}$. Therefore, $f_{syn}$ is updated depending on datasets $\{w_{\alpha},x_{\alpha}\}_{\alpha=1}^{6}$ in the synchronization phase as follows,

\begin{align}
f_{syn}=\{w_{1},x_{1}\}{\oplus}\{w_{2},x_{2}\}{\oplus}\{w_{3},x_{3}\}{\,\cdots\,}{\oplus}\{w_{6},x_{6}\}
\end{align}

{\noindent}where the operator ${\oplus}$ means all-to-all connections between the oscillation units.

In the network in Fig. \ref{fig. 1}(b), each unit can operate at a resonant frequency with a micro/nano-electromechanical systems (MEMS/NEMS) resonator \cite{Cleland1996NEMS}\text{\color{mygreen}--}\cite{Xu2022NEMS}, and datasets $\{w_{\alpha},x_{\alpha}\}_{\alpha=1}^{6}$ can shift the resonant frequencies of resonators, thereby updating $f_{syn}$. Furthermore, by leveraging the NVM characteristics of resonators \cite{Badzey2004NVM}\text{\color{mygreen}--}\cite{Uka2025NVM}, a frequency-domain AIMC can be implemented during the synchronization phase as depicted in Fig. \ref{fig. 1}(b). 
\section{\color{myblue}\Large{F}\large{REQUENCY} \Large{S}\large{YNCHRONIZATION}}
\begin{figure}[h!]
\centering\includegraphics[scale = 0.55]{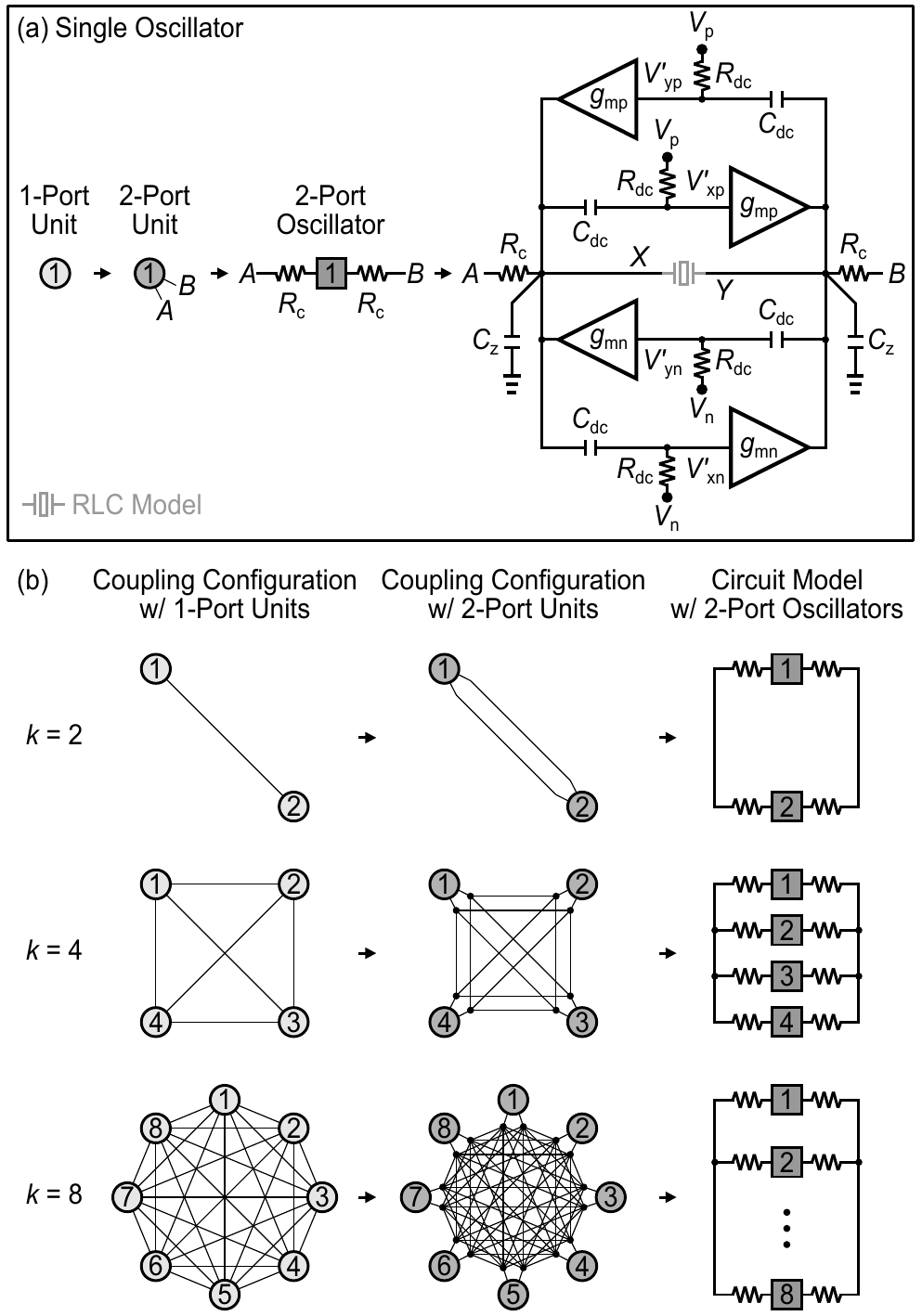}
\captionof{figure}{{\color{myred}\textbf{Coupling configuration.}} (a) Single unit. (b) Configuration depending on the number of coupled oscillators.}\label{fig. 2}
\end{figure}

To implement the synchronization model shown in Fig. \ref{fig. 1}(b), the single oscillator depicted in Fig. \ref{fig. 2}(a) is employed. To simplify the analysis of the all-to-all connections in Fig. \ref{fig. 1}(b), the single oscillator is first expressed by adopting a \textit{1-Port Unit} in Fig. \ref{fig. 2}(a). Then, since the coupling configuration is ultimately implemented using a \textit{2-Port Oscillator}, the \textit{1-Port Unit} is extended to a \textit{2-Port Unit}. Finally, the \textit{2-Port Unit} is expressed equivalently as the \textit{2-Port Oscillator}. The \textit{1-Port Unit}, \textit{2-Port Unit}, and \textit{2-Port Oscillator} are all identical, but employed to effectively explain the extension from a widely used coupling network (consisting of \textit{1-Port Unit}s) to a proposed oscillator-based coupling network.

Fig. \ref{fig. 2}(b) shows the coupling configurations depending on the number of coupled oscillators, $k$, using \textit{1-Port Unit}s, \textit{2-Port Unit}s, and \textit{2-Port Oscillator}s. The coupling configuration using \textit{1-Port Unit}s has been widely adopted as an oscillatory neural network (ONN) \cite{Todri-Sanial2024ONN} for combinatorial optimization \cite{Mohseni2022Ising}\text{\color{mygreen}--}\cite{Wang2019Ising}, pattern recognition \cite{Hoppensteadt2000pattern}\text{\color{mygreen}--}\cite{Shukla2016pattern}, and image processing \cite{Kim2023edge}, \cite{Delacour2024edge}. In this work, as the frequency synchronization is implemented based on a 2-port circuit model, the coupling configuration evolves from the \textit{1-Port Unit} to the \textit{2-Port Unit}, and subsequently to the \textit{2-Port Oscillator}, as depicted in Fig. \ref{fig. 2}. The resistor $R_{c}$ connected across the \textit{2-Port Oscillator} determines the coupling strength between the units. A lower value of $R_{c}$ improves the coupling strength between the units, thereby effectively synchronizing the individual frequencies to $f_{syn}$.

\subsection{\color{myblue}Single Unit}
The \textit{2-Port Oscillator} is designed using transconductance elements (two $g_{mp}$ and two $g_{mn}$), high-pass filters (four $R_{dc}$ and four $C_{dc}$), coupling resistors (two $R_{c}$), and load capacitors (two $C_{z}$) as depicted in Fig. \ref{fig. 2}(a). In the oscillator configuration shown in Fig. \ref{fig. 2}(a), the upper half forms a cross-coupled pair consisting of $g_{mp}$, $R_{dc}$, and $C_{dc}$. Similarly, the lower half also forms a cross-coupled pair consisting of $g_{mn}$, $R_{dc}$, and $C_{dc}$. Since each cross-coupled pair has a resonant frequency formed by a high-pass filter ($R_{dc}$ and $C_{dc}$) and a low-pass filter ($g_{mp}$ or $g_{mn}$), and sustains positive feedback, each pair oscillates at that frequency even without a mechanical resonator.

The complementary configuration{\textemdash}consisting of the upper and lower halves{\textemdash}can determine an oscillation frequency with the control of two voltage signals, $V_{p}$ and $V_{n}$. In Fig. \ref{fig. 2}(a), $V_{p}$ and $V_{n}$ contribute to a resonant frequency by tuning $g_{mp}$ and $g_{mn}$, respectively. In addition, the variations in $R_{dc}$, $C_{dc}$, and $C_{z}$ also contribute to the resonant frequency. While the RLC model between nodes $X$ and $Y$ represents a MEMS/NEMS resonator that contributes to the resonant frequency, the \textit{2-Port Oscillator} can oscillate even without the resonator, as the upper and lower cross-coupled pairs generate a resonance peak and sustain positive feedback.

Note that $V_{p}$ and $V_{n}$ must be set through a digital-to-analog converter based on dataset $\{w_{\alpha},x_{\alpha}\}$ depicted in Fig. \ref{fig. 1}(b) so that the multiplication of $w_{\alpha}$ and $x_{\alpha}$ can appropriately represent the resonant frequency.

\subsection{\color{myblue}Small-Signal Model of a Single Unit}
\begin{figure}[h!]
\centering\includegraphics[scale = 0.55]{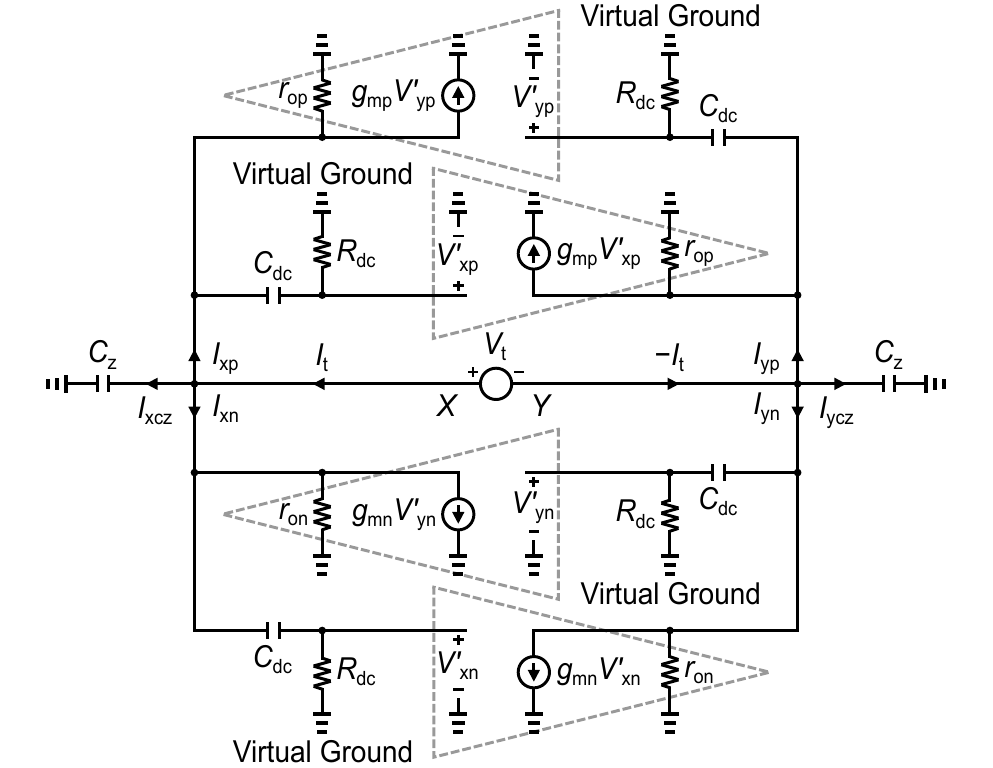}
\captionof{figure}{{\color{myred}\textbf{Small-signal model of the \textit{2-Port Oscillator}.}} Complementary configuration consisting of the upper ($p$-type) and lower ($n$-type) cross-coupled pairs.}\label{fig. 3}
\end{figure}

To understand the behavior of a resonant frequency, the small-signal model of the \textit{2-Port Oscillator} is drawn as shown in Fig. \ref{fig. 3}. \footnote{$V_{p}$ and $V_{n}$ shown in Fig. \ref{fig. 2}(a) become virtual grounds in the small-signal model. For simplicity, the coupling resistor $R_{c}$ is excluded from the small-signal analysis.} In Fig. \ref{fig. 3}, each transconductance element is depicted as an isosceles triangle using a gray dotted line. $r_{op}$ and $r_{on}$ represent the output resistances for the upper ($p$-type) and lower ($n$-type) transconductance elements, respectively. To understand the impedance across nodes $X$ and $Y$ in the frequency domain, applying a test voltage $V_{t}$ and solving Kirchhoff's Current Law (KCL) across nodes $X$ and $Y$ in Fig. \ref{fig. 3}, the resulting currents, $I_{t}$ and $-I_{t}$, are drawn as follows:

\begin{align}
I_{t}&=I_{xp}+I_{xn}+I_{xcz}\\
-I_{t}&=I_{yp}+I_{yn}+I_{ycz}
\end{align}

{\noindent}where $I_{xp}$ and $I_{yp}$ are the small-signal currents drawn by the upper cross-coupled pair, $I_{xn}$ and $I_{yn}$ are the small-signal currents drawn by the lower cross-coupled pair, and $I_{xcz}$ and $I_{ycz}$ flow through $C_{z}$ from nodes $X$ and $Y$, respectively.

Assuming that $r_{op}$ and $r_{on}$ are sufficiently large so that the small-signal currents flowing through $r_{op}$ and $r_{on}$ can be negligible, $I_{t}$ and $-I_{t}$ in Equations (2) and (3) can be expanded as

\begin{align}
I_{t}&=\underbrace{g_{mp}V'_{yp}+\frac{V_{x}}{Z_{dc}}}_{I_{xp}}+\underbrace{g_{mn}V'_{yn}+\frac{V_{x}}{Z_{dc}}}_{I_{xn}}+\underbrace{sC_{z}V_{x}}_{I_{xcz}}\\
-I_{t}&=\underbrace{g_{mp}V'_{xp}+\frac{V_{y}}{Z_{dc}}}_{I_{yp}}+\underbrace{g_{mn}V'_{xn}+\frac{V_{y}}{Z_{dc}}}_{I_{yn}}+\underbrace{sC_{z}V_{y}}_{I_{ycz}}
\end{align}

{\noindent}where $V_{x}$ and $V_{y}$ are respectively the voltages at nodes $X$ and $Y$, $V'_{xp}=V'_{xn}=sR_{dc}C_{dc}V_{x}/(1+sR_{dc}C_{dc})$, $V'_{yp}=V'_{yn}=sR_{dc}C_{dc}V_{y}/(1+sR_{dc}C_{dc})$, and $Z_{dc}=(1+sR_{dc}C_{dc})/sC_{dc}$. Then, subtracting the two currents, $I_{t}$ and $-I_{t}$, and using $V_{t}=V_{x}-V_{y}$, the impedance seen looking into nodes $X$ and $Y$ is given by

\begin{align}
\frac{V_{t}}{I_{t}}&=\frac{-2}{G_{m}}+j{\omega}\frac{-2C_{z}}{G_{m}^2}+\frac{2}{-j{\omega}G_{m}R_{dc}C_{dc}}\\
&=R_{t}+j{\omega}L_{n}+\frac{1}{j{\omega}C_{n}}
\end{align}

{\noindent}where $G_{m}=g_{mp}+g_{mn}$, $R_{t}=-2/G_{m}$, $L_{n}=-2C_{z}/(G_{m}^2)$, and $C_{n}=-G_{m}R_{dc}C_{dc}/2$. \footnote{Details of the small-signal analysis for the \textit{2-Port Oscillator} are provided in Appendix A.1, which is an expanded version of \cite{Lee2026osc}.} The impedance $V_{t}/I_{t}$ consists of a negative resistance $R_{t}$, a negative inductance $L_{n}$, and a negative capacitance $C_{n}$ connected in series. Therefore, the \textit{2-Port Oscillator} can operate as an active resonator.

In this series RLC circuit{\textemdash}$R_{t}$, $L_{n}$, and $C_{n}$ from Equation (7), resonance occurs at a frequency of $1/(2{\pi}\sqrt{L_{n}C_{n}})$. Since $R_{t}$ is negative, energy at $1/(2{\pi}\sqrt{L_{n}C_{n}})$ is sustained in the loop of the series RLC circuit, unlike the case where energy is dissipated via a positive resistance, thereby generating oscillation at $1/(2{\pi}\sqrt{L_{n}C_{n}})$. Note that while a positive resistance absorbs energy from remaining circuits, a negative resistance delivers energy to the remaining circuits \cite{Chua1987resistor}. Consequently, the resonant frequency of the \textit{2-Port Oscillator} is given by

\begin{align}
f&=\frac{1}{2{\pi}\sqrt{L_{n}C_{n}}}\\
&=\frac{1}{2{\pi}}\sqrt{\frac{G_{m}}{C_{z}R_{dc}C_{dc}}}\\
&=\frac{1}{2{\pi}}\frac{1}{\sqrt{R_{Gm}C_{z}R_{dc}C_{dc}}}
\end{align}

{\noindent}where $G_{m}$ is expressed as $1/R_{Gm}$. Even though the circuit model in Fig. \ref{fig. 3} is not connected to a mechanical resonator, it oscillates at a frequency of $1/(2{\pi}\sqrt{R_{Gm}C_{z}R_{dc}C_{dc}})$, which is established through the resonance peak, positive feedback, and negative resistance of the cross-coupled pairs.

Recall that the dataset $\{w_{1},x_{1}\}$ in Fig. \ref{fig. 1}(b) is encoded for a frequency of $f_{1}$ using an oscillation unit, the encoding process{\textemdash}frequency domain multiplication{\textemdash}is conducted through the following functions:

\begin{align}
\underbrace{\mathcal{F}\left(\underbrace{\mathcal{M}\left(\underbrace{\mathcal{N}\left(w_{1},x_{1}\right)}_{V_{p[1]},V_{n[1]}}\right)}_{G_{m[1]}}\right)}_{f_{1}}
\end{align}

{\noindent}where the function $\mathcal{N}(\cdot)$ takes $w_{1}$ and $x_{1}$ as inputs and returns the two voltage signals, $V_{p[1]}$ and $V_{n[1]}$, corresponding to $V_{p}$ and $V_{n}$ in Fig. \ref{fig. 2}(a), respectively; the function $\mathcal{M}(\cdot)$ takes $V_{p[1]}$ and $V_{n[1]}$ as inputs and returns $G_{m[1]}$, corresponding to $G_{m}=g_{mp}+g_{mn}$ in Equation (9); and the function $\mathcal{F}(\cdot)$ returns $f_{1}$ by taking $G_{m[1]}$ as input. Therefore, from Equation (11), the signal flow for the encoding can be rewritten as

\begin{align}
\left\{w_{1},x_{1}\right\}\xrightarrow{\mathcal{N}(\cdot)}\left\{V_{p[1]},V_{n[1]}\right\}\xrightarrow{\mathcal{M}(\cdot)}G_{m[1]}\xrightarrow{\mathcal{F}(\cdot)}f_{1}
\end{align}

{\noindent}Note that the function $\mathcal{N}(\cdot)$ requires the nonlinear conversion of $\{w_{1},x_{1}\}$ into $\{V_{p[1]},V_{n[1]}\}$ using a digital-to-analog converter for the process of $w_{1}x_{1}{\rightarrow}G_{m[1]}$. However, this work does not address the implementation of $\mathcal{N}(\cdot)$ and focuses solely on $\mathcal{M}(\cdot)$ and $\mathcal{F}(\cdot)$ using the \textit{2-Port Oscillator}. 
\subsection{\color{myblue}Frequency in Coupled Oscillators}
\begin{figure}[h!]
\centering\includegraphics[scale = 0.55]{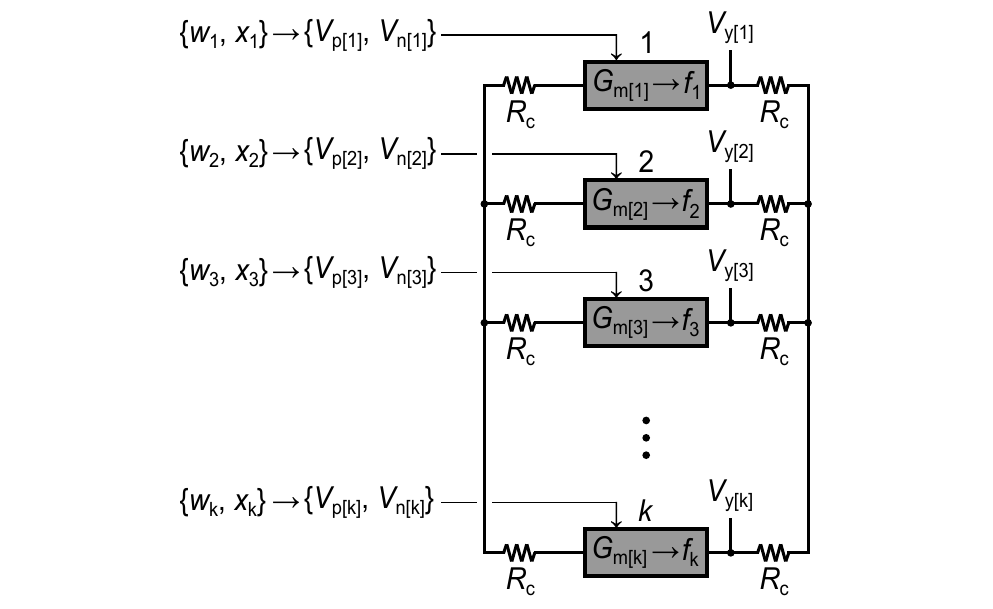}
\captionof{figure}{{\color{myred}\textbf{Coupled oscillators.}} Frequency synchronization model using \textit{2-Port Oscillator}s indicated by gray boxes.}\label{fig. 4}
\end{figure}

A frequency synchronization model is implemented by expanding the \textit{2-Port Oscillator} as shown in Fig. \ref{fig. 4}, where $k$ is the number of coupled \textit{2-Port Oscillator}s. Recall that the single oscillator in Fig. \ref{fig. 2}(a) is expanded into a network where all units are connected in an all-to-all manner as shown in Fig. \ref{fig. 2}(b), the $p$-type voltages $V_{p[1]}$ through $V_{p[k]}$ control the transconductances $g_{mp[1]}$ through $g_{mp[k]}$ of the $p$-type cross-coupled pairs, the $n$-type voltages $V_{n[1]}$ through $V_{n[k]}$ control the transconductances $g_{mn[1]}$ through $g_{mn[k]}$ of the $n$-type cross-coupled pairs, and $G_{m[1]}$ through $G_{m[k]}$ are respectively expressed as $g_{mp[1]}+g_{mn[1]}$ through $g_{mp[k]}+g_{mn[k]}$ in Fig. \ref{fig. 4}. The node voltages $V_{y[1]}$ through $V_{y[k]}$ correspond to the voltage at node $Y$ of the \textit{2-Port Oscillator} in Fig. \ref{fig. 2}(a).

In Fig. \ref{fig. 4}, assuming that $R_{c}$ is sufficiently low to connect all units and applying a test voltage $V_{t}$ across the units, the small-signal currents, $I_{t}$ and $-I_{t}$, flowing through the units are scaled by a factor of $k$ using Equations (2) and (3), as follows. Note that $I_{t}$ and $-I_{t}$ flow through the left and right sides of the coupled oscillators in Fig. \ref{fig. 4}, respectively.

\begin{align}
I_{t}&=k\left(I_{xp}+I_{xn}+I_{xcz}\right)\\
-I_{t}&=k\left(I_{yp}+I_{yn}+I_{ycz}\right)
\end{align}

{\noindent}where the currents{\textemdash}$I_{xp}$, $I_{xn}$, $I_{xcz}$, $I_{yp}$, $I_{yn}$, and $I_{ycz}${\textemdash}are the same as the currents of the small-signal model depicted in Fig. \ref{fig. 3}. When scaling a single oscillator by a factor of $k$ using the small-signal model of the \textit{2-Port Oscillator} (Fig. \ref{fig. 3}), $I_{t}$ and $-I_{t}$ in Equations (13) and (14) are expressed as

\begin{align}
I_{t}=&\underbrace{\sum_{\alpha=1}^{k}g_{mp[\alpha]}V'_{yp}+\frac{kV_{x}}{Z_{dc}}}_{k{\cdot}I_{xp}}+\underbrace{\sum_{\alpha=1}^{k}g_{mn[\alpha]}V'_{yn}+\frac{kV_{x}}{Z_{dc}}}_{k{\cdot}I_{xn}}\\
{\nonumber}&+\underbrace{sC_{z}kV_{x}}_{k{\cdot}I_{xcz}}\\
-I_{t}=&\underbrace{\sum_{\alpha=1}^{k}g_{mp[\alpha]}V'_{xp}+\frac{kV_{y}}{Z_{dc}}}_{k{\cdot}I_{yp}}+\underbrace{\sum_{\alpha=1}^{k}g_{mn[\alpha]}V'_{xn}+\frac{kV_{y}}{Z_{dc}}}_{k{\cdot}I_{yn}}\\
{\nonumber}&+\underbrace{sC_{z}kV_{y}}_{k{\cdot}I_{ycz}}
\end{align}

{\noindent}where $V'_{xp}$, $V'_{xn}$, $V'_{yp}$, $V'_{yn}$, and $Z_{dc}$ are the same as those used in Equations (4) and (5). $V_{x}$ and $V_{y}$ are respectively the node voltages at the left and right sides of the coupled oscillators in Fig. \ref{fig. 4} when $R_{c}$ is sufficiently low.

Subtracting the above two currents, $I_{t}$ and $-I_{t}$, and applying $V_{t}=V_{x}-V_{y}$ to that result, the impedance across the coupled oscillators is given by

\begin{align}
\frac{V_{t}}{I_{t}}=&\frac{-2}{\sum_{\alpha=1}^{k}G_{m[\alpha]}}+j{\omega}\frac{-2C_{z}k}{\left(\sum_{\alpha=1}^{k}G_{m[\alpha]}\right)^2}\\
{\nonumber}&+\frac{2}{-j{\omega}\sum_{\alpha=1}^{k}G_{m[\alpha]}R_{dc}C_{dc}}
\end{align}

{\noindent}where $G_{m[\alpha]}$ is defined as $g_{mp[\alpha]}+g_{mn[\alpha]}$, representing total transconductance of $p$-type and $n$-type in each oscillator. Therefore, similar to Equations (6) and (7), $V_{t}/I_{t}$ of the coupled oscillators can be expressed as a series RLC circuit consisting of a negative resistance $R_{tc}$, a negative inductance $L_{nc}$, and a negative capacitance $C_{nc}$, as follows:

\begin{align}
R_{tc}+j{\omega}L_{nc}+\frac{1}{j{\omega}C_{nc}}
\end{align}

{\noindent}where $R_{tc}$, $L_{nc}$, and $C_{nc}$ are expressed using the scaled transconductance depending on the number of coupled oscillators. \footnote{In the coupled oscillators in Fig. \ref{fig. 4}, details of the small-signal analysis for $V_{t}/I_{t}$, $R_{tc}$, $L_{nc}$, and $C_{nc}$ are provided in Appendix A.2.} Each element is summarized as follows:

\begin{align}
R_{tc}&=\frac{-2}{\sum_{\alpha=1}^{k}G_{m[\alpha]}}\\
L_{nc}&=\frac{-2C_{z}k}{\left(\sum_{\alpha=1}^{k}G_{m[\alpha]}\right)^2}\\
C_{nc}&=\frac{-\sum_{\alpha=1}^{k}G_{m[\alpha]}R_{dc}C_{dc}}{2}
\end{align}

Similar to the resonance of the \textit{2-Port Oscillator}, the series RLC circuit derived from the coupled oscillators also generates resonance at a frequency of $1/(2{\pi}\sqrt{L_{nc}C_{nc}})$, which is defined as the synchronized frequency $f_{syn}$. In the loop of the series RLC circuit, the negative resistance $R_{tc}$ supplies energy to $L_{nc}$ and $C_{nc}$, thereby generating $f_{syn}$ across the coupled oscillators depicted in Fig. \ref{fig. 4} as follows:

\begin{align}
f_{syn}&=\frac{1}{2{\pi}\sqrt{L_{nc}C_{nc}}}\\
&=\frac{1}{2{\pi}}\sqrt{\frac{\sum_{\alpha=1}^{k}G_{m[\alpha]}}{C_{z}kR_{dc}C_{dc}}}\\
&=\frac{1}{2{\pi}}\sqrt{\sum_{\alpha=1}^{k}\frac{1}{R_{Gm[\alpha]}C_{z}kR_{dc}C_{dc}}}
\end{align}

{\noindent}where $G_{m[\alpha]}=g_{mp[\alpha]}+g_{mn[\alpha]}=1/R_{Gm[\alpha]}$. Note that $R_{tc}$, $L_{nc}$, $C_{nc}$, and $f_{syn}$ are obtained assuming that $R_{c}$ in Fig. \ref{fig. 4} is sufficiently low to ensure the coupling of all oscillation units. Therefore, when $R_{c}$ is high enough in Fig. \ref{fig. 4}, the oscillators are not synchronized and exhibit independent frequencies such as $f_{1}$, $f_{2}{\cdots}f_{k}$. Then, as $R_{c}$ becomes low enough, the oscillators are synchronized, and their frequencies are updated as follows: $f_{1}{\rightarrow}f_{syn}$, $f_{2}{\rightarrow}f_{syn}{\cdots}f_{k}{\rightarrow}f_{syn}$.

Recall that when the datasets $\{w_{1},x_{1}\}$ through $\{w_{k},x_{k}\}$ in Fig. \ref{fig. 4} are applied to the coupled oscillators in the synchronization mode, the signal flow is expressed as

\begin{align}
\begin{gathered}
\left\{w_{\alpha},x_{\alpha}\right\}_{\alpha=1}^{k}\\
\big\downarrow{\,\mathcal{N}_{c}(\cdot)}\\
\left\{V_{p[\alpha]},V_{n[\alpha]}\right\}_{\alpha=1}^{k}\\
\big\downarrow{\,\mathcal{M}_{c}(\cdot)}\\
\left\{G_{m[\alpha]}\right\}_{\alpha=1}^{k}\\
\big\downarrow{\,\mathcal{F}_{c}(\cdot)}\\
\boldsymbol{f_{syn}}
\end{gathered}
\end{align}

{\noindent}where the function $\mathcal{N}_{c}(\cdot)$ takes $\{w_{1},x_{1}\}$ through $\{w_{k},x_{k}\}$ and returns $\{V_{p[1]},V_{n[1]}\}$ through $\{V_{p[k]},V_{n[k]}\}$, the function $\mathcal{M}_{c}(\cdot)$ takes $\{V_{p[1]},V_{n[1]}\}$ through $\{V_{p[k]},V_{n[k]}\}$ and returns $G_{m[1]}$ through $G_{m[k]}$, and the function $\mathcal{F}_{c}(\cdot)$ generates $f_{syn}$ combining $G_{m[1]}$ through $G_{m[k]}$.

\begin{figure}[h!]
\centering\includegraphics[scale = 0.55]{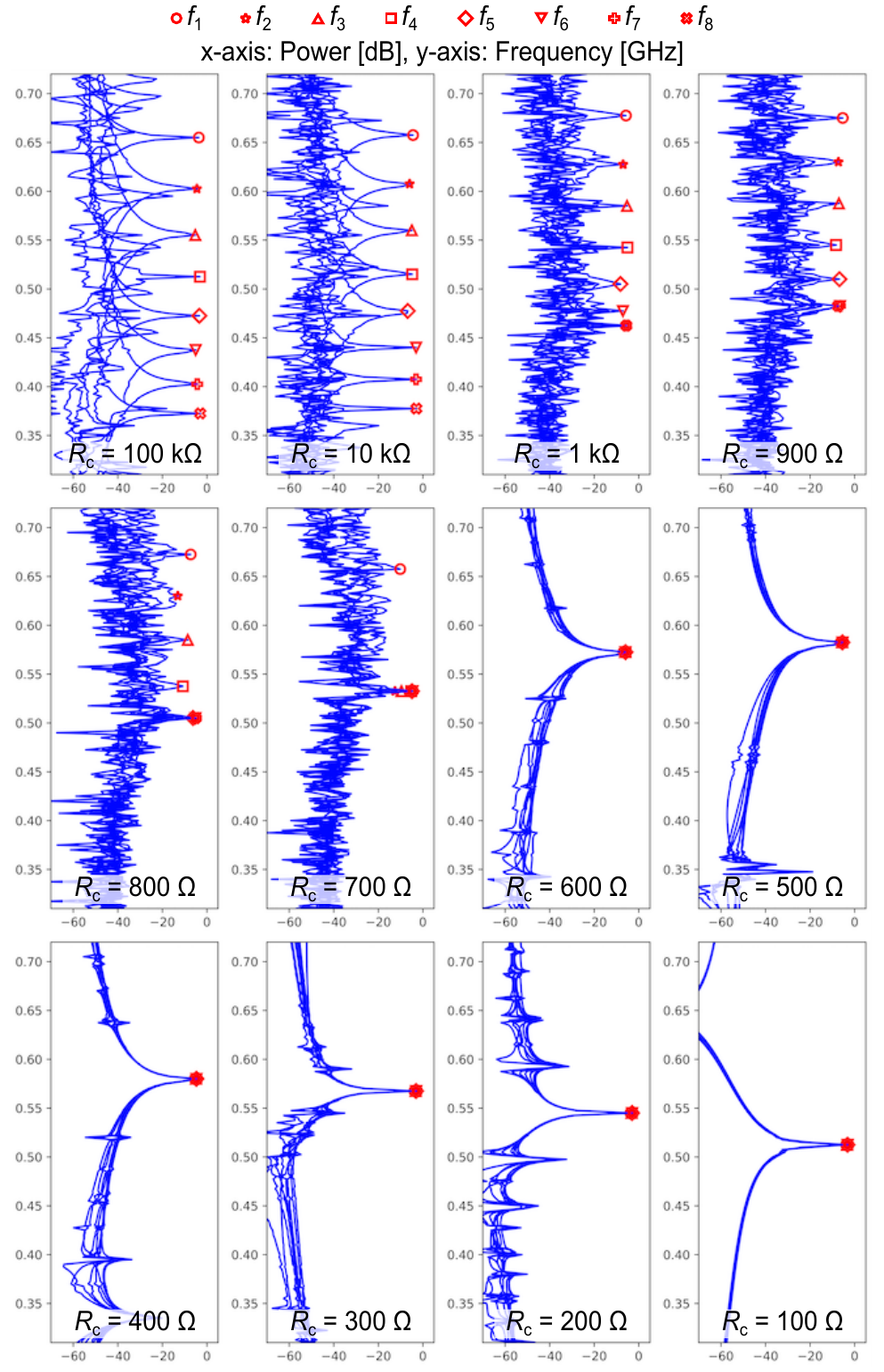}
\captionof{figure}{{\color{myred}\textbf{Frequency synchronization.}} Frequency variations of the respective oscillators in $k=8$ when the coupling resistor $R_{c}$ changes from $100{\,}\mathrm{k}{\Omega}$ to $100{\,}{\Omega}$ (x-axis and y-axis indicate $\mathrm{Power{\,}[dB]}$ and $\mathrm{Frequency{\,}[GHz]}$, respectively). In a $3{\times}4$ layout, the coupling strength increases along the directions of left-to-right and top-to-bottom.}\label{fig. 5}
\end{figure}

Fig. \ref{fig. 5} presents the process by which the frequencies of individual oscillators become synchronized in $k=8$ when the coupling strength changes by controlling $R_{c}$. \footnote{Details of the circuit implementation for the frequency synchronization are provided in Appendix A.3. Fig. \ref{fig. 5} is obtained based on the model depicted in Fig. \ref{fig. 4} by using Cadence Virtuoso Studio (IC23.1-64b.43).} In 12 subplots arranged in a $3\times4$ layout (Fig. \ref{fig. 5}), the peak frequencies $f_{1}$ through $f_{8}$ observed in the individual oscillators depicted in Fig. \ref{fig. 4} are denoted using different red symbols. $f_{1}$ through $f_{8}$ are the results of the fast Fourier transform (FFT) performed using $V_{y[1]}$ through $V_{y[8]}$ shown in Fig. \ref{fig. 4}.

As $R_{c}$ that determines the coupling strength changes from $100{\,}\mathrm{k}{\Omega}$ to $900{\,}{\Omega}$ (top row), those in the relatively low-frequency region begin to synchronize first. Then, as $R_{c}$ changes from $800{\,}{\Omega}$ to $500{\,}{\Omega}$ (middle row), those in the relatively high-frequency region also move toward the already synchronized frequencies. Immediately after all frequencies are synchronized at $R_{c}=600{\,}{\Omega}$, spectral regrowth and harmonics appear around the synchronized frequencies, but these can be mitigated by improving the coupling strength (bottom row). When $R_{c}$ reaches $100{\,}{\Omega}$, the oscillators exhibit improved synchronization with mitigated spectral regrowth and harmonics, compared to the cases where $R_{c}$ changes from $600{\,}{\Omega}$ to $200{\,}{\Omega}$ (Fig. \ref{fig. 5}). As a result, when the oscillators are not synchronized ($R_{c}=100{\,}\mathrm{k}{\Omega}$), the frequencies $f_{1}$ through $f_{8}$ are distributed from $0.372{\,}\mathrm{GHz}$ to $0.655{\,}\mathrm{GHz}$ with an average frequency interval of $0.04{\,}\mathrm{GHz}$, then when $R_{c}$ reaches $100{\,}{\Omega}$, the individual frequencies are synchronized to $0.512{\,}\mathrm{GHz}$, which corresponds to $f_{syn}$.

Therefore, as the coupling strength increases by reducing $R_{c}$, the individual frequencies $f_{1}$ through $f_{8}$ gradually shift to $f_{syn}$ with improved synchronization characteristics. \footnote{To set the individual frequencies $f_{1}$ through $f_{8}$ in Fig. \ref{fig. 5}, the voltages $\{V_{p[1]},V_{n[1]}\}$ through $\{V_{p[8]},V_{n[8]}\}$ are applied to the circuit model, and the used voltages are provided in Appendix A.3.} Note that the intervals between the individual frequencies depend on the design parameters: $\{V_{p[\alpha]},V_{n[\alpha]}\}_{\alpha=1}^{k}$, number of coupled oscillators $k$, $R_{dc}$, $C_{dc}$, and $C_{z}$. Therefore, the value of $R_{c}$ that enables complete synchronization changes according to these design parameters.

\begin{figure}[h!]
\centering\includegraphics[scale = 0.55]{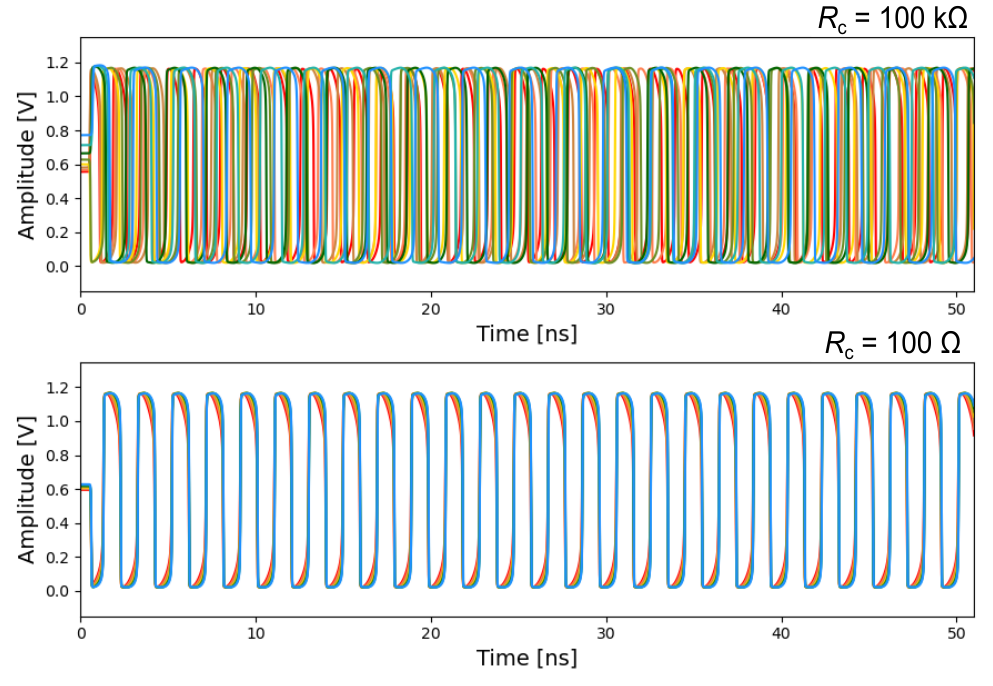}
\captionof{figure}{{\color{myred}\textbf{Transient response.}} Synchronization of the respective oscillators at $R_{c}=100{\,}\mathrm{k}{\Omega}$ and $R_{c}=100{\,}{\Omega}$.}\label{fig. 6}
\end{figure}

Fig. \ref{fig. 6} shows the transient response of the coupled oscillators in $k=8$ depending on the different coupling strengths. When $R_{c}=100{\,}\mathrm{k}{\Omega}$, the individual oscillators are not synchronized in both frequency and phase, as shown in the top of Fig. \ref{fig. 6}. Then, as $R_{c}$ is reduced to $100{\,}{\Omega}$, the oscillators are synchronized in both frequency and phase, as shown in the bottom of Fig. \ref{fig. 6}. The FFT results of Fig. \ref{fig. 6} for $R_{c}=100{\,}\mathrm{k}{\Omega}$ and $R_{c}=100{\,}{\Omega}$ are presented in Fig. \ref{fig. 5}. 
\section{\color{myblue}\Large{C}\large{OUPLING} \large{WITH} \Large{R}\large{ESONATORS}}
In Section $\mathrm{II}$, the analysis of the frequency synchronization is conducted without the resonator model depicted in Fig. \ref{fig. 2}(a). This section covers the frequency synchronization circuit model, including the resonator model. A mechanical resonator can be developed using MEMS or NEMS fabrication technology, and the modified Butterworth-Van Dyke (mBVD) model is widely adopted as the equivalent circuit model for the resonator \cite{Larson1999mBVD}\text{\color{mygreen}--}\cite{Lin2014mBVD}. In this work, the mBVD model is transformed into a parallel RLC model to facilitate the analysis of the small-signal circuit model.

\subsection{\color{myblue}Resonator Model Approximation}
\begin{figure}[h!]
\centering\includegraphics[scale = 0.55]{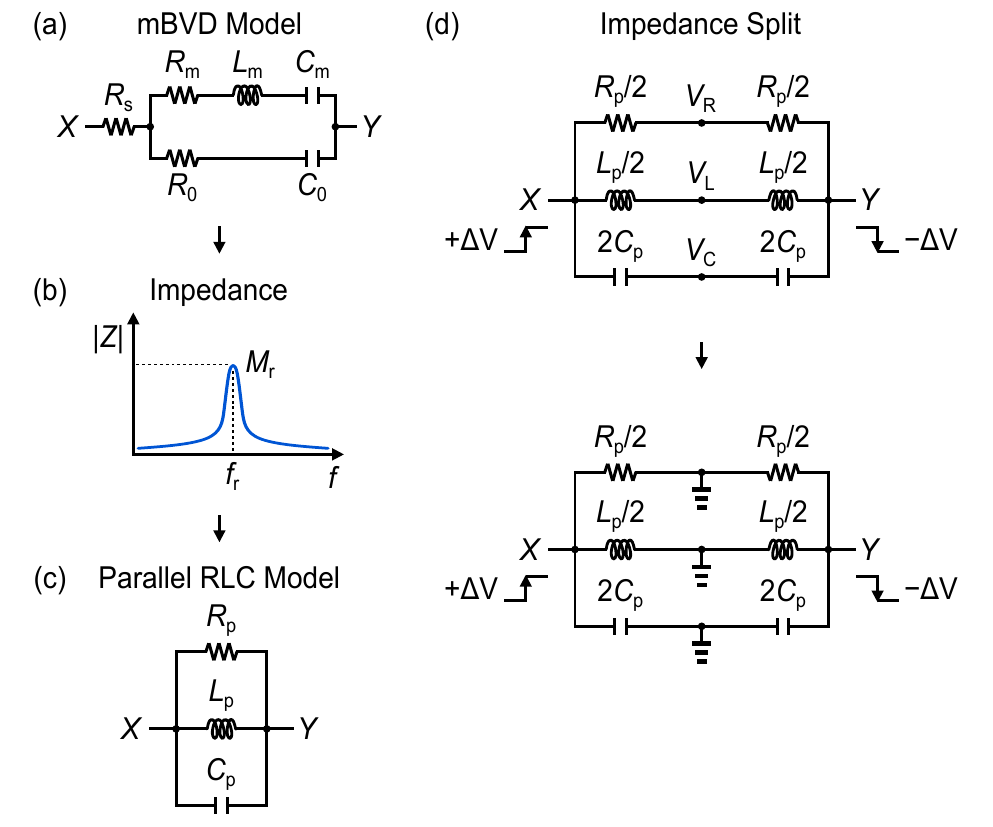}
\captionof{figure}{{\color{myred}\textbf{Impedance transformation.}} (a) mBVD model. (b) Impedance of the mBVD model. (c) Equivalent parallel RLC model. (d) Split of the equivalent parallel RLC model.}\label{fig. 7}
\end{figure}

The mBVD model in Fig. \ref{fig. 7}(a) consists of motional elements ($R_{m}$, $L_{m}$, and $C_{m}$), static elements ($R_{0}$ and $C_{0}$), and a series electrode resistance $R_{s}$ \cite{Larson1999mBVD}\text{\color{mygreen}--}\cite{Lin2014mBVD}. The impedance of the mBVD model is described in Fig. \ref{fig. 7}(b), and its resonant frequency $f_{r}$ is formed at $1/(2{\pi}\sqrt{L_{m}C_{m}})$ with a peak magnitude of $M_{r}$. To mimic the impedance in Fig. \ref{fig. 7}(b) using a simplified circuit model, a parallel RLC model is employed with the elements $R_{p}$, $L_{p}$, and $C_{p}$ depicted in Fig. \ref{fig. 7}(c). \footnote{In the parallel RLC circuit consisting of $R_{p}$, $L_{p}$, and $C_{p}$, its resonant frequency is achieved as $1/(2{\pi}\sqrt{L_{p}C_{p}})$ with a peak magnitude of $R_{p}$.} Since the resonant frequency of the parallel RLC model must imitate the resonant frequency of the mBVD model, $L_{p}$ and $C_{p}$ must be equal to $L_{m}$ and $C_{m}$, respectively.

Given that a series resistance dominates the impedance magnitude at the resonant frequency of a series RLC circuit, whereas a parallel resistance dominates the impedance magnitude at the resonant frequency of a parallel RLC circuit, $R_{s}$ of the mBVD model can be correlated with $R_{p}$ of the parallel RLC model. Therefore, in this work, the mBVD-to-parallel RLC transformation is conducted by using the assumptions of $L_{p}=L_{m}$, $C_{p}=C_{m}$, and $R_{p}=M_{r}=k_{t}/R_{s}$, where $M_{r}$ is the peak magnitude in Fig. \ref{fig. 7}(b), and $k_{t}$ is the transformation coefficient. \footnote{$k_{t}$ is adopted to map the peak magnitude of the mBVD model to that of the parallel RLC model, and $k_{t}$ depends on the resonator design.}

Applying the parallel RLC model, which is approximated from the mBVD model, to the \textit{2-Port Oscillator} in Fig. \ref{fig. 2}(a), each parallel element can be split into two series elements as depicted in the top of Fig. \ref{fig. 7}(d). For simplicity of the small-signal analysis, the parallel RLC model is split into two parallel structures as depicted in the top of Fig. \ref{fig. 7}(d). Since the circuit configuration of the \textit{2-Port Oscillator} is fully differential, the signals at nodes $X$ and $Y$ change by the same amount, but in opposite directions. Therefore, when the signal at node $X$ changes by $+{\Delta}V$, the signal at node $Y$ changes by $-{\Delta}V$. To find the midpoint potentials ($V_{R}$, $V_{L}$, and $V_{C}$) in the top of Fig. \ref{fig. 7}(d), the current $I_{R}$ flowing through two $R_{p}/2$ is given by $[+{\Delta}V-(-{\Delta}V)]/R_{p}=2{\Delta}V/R_{p}$, then the voltage drop of each $R_{p}/2$ is obtained as $I_{R}(R_{p}/2)={\Delta}V$ \cite{Razavi2017circuit}. Therefore, the midpoint potential $V_{R}$ is a virtual (or ac) ground. Similarly, $V_{L}$ and $V_{C}$ are also virtual grounds, respectively. Consequently, as shown in the bottom of Fig. \ref{fig. 7}(d), the split parallel RLC model with a virtual ground at each midpoint is used with the small-signal model of the \textit{2-Port Oscillator} to analyze the frequency synchronization involving resonators. 
\subsection{\color{myblue}Single Unit Involving a Resonator Model}
\begin{figure}[h!]
\centering\includegraphics[scale = 0.55]{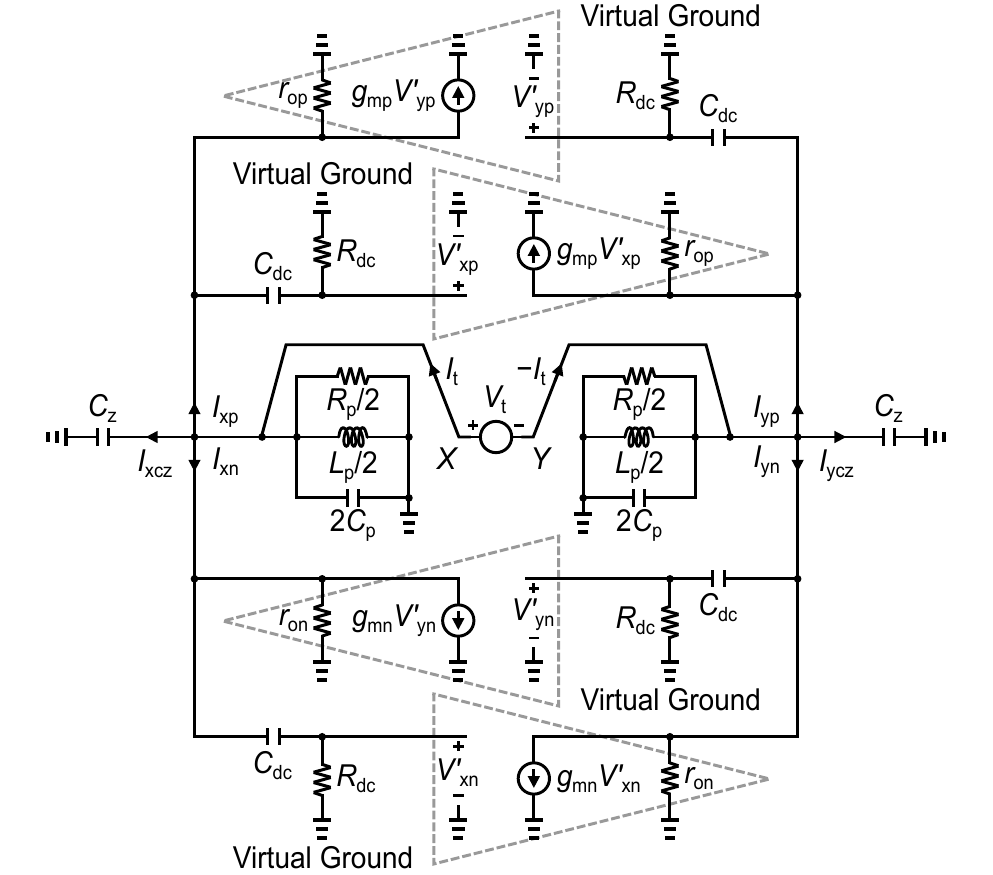}
\captionof{figure}{{\color{myred}\textbf{Small-signal model of the \textit{2-Port Oscillator} involving a resonator model.}} Complementary configuration consisting of the upper ($p$-type) and lower ($n$-type) cross-coupled pairs with the split parallel RLC model.}\label{fig.8}
\end{figure}

Fig. \ref{fig.8} presents the small-signal model of the \textit{2-Port Oscillator} from Fig. \ref{fig. 2}(a), incorporating a resonator model approximated by a split parallel RLC model. As with the small-signal model in Fig. \ref{fig. 3}, $V_{p}$ and $V_{n}$ depicted in Fig. \ref{fig. 2}(a) become virtual grounds in Fig. \ref{fig.8}. The equations for $V'_{xp}$, $V'_{xn}$, $V'_{yp}$, $V'_{yn}$, and $Z_{dc}$ are the same as those used in the analysis of Fig. \ref{fig. 3}. In Fig. \ref{fig.8}, the isosceles triangles drawn with gray dotted lines represent the transconductance elements shown in Fig. \ref{fig. 2}(a).

In Fig. \ref{fig.8}, assuming that $r_{op}$ and $r_{on}$ are large enough so that the currents flowing through $r_{op}$ and $r_{on}$ can be negligible, and applying a test voltage $V_{t}$ between nodes $X$ and $Y$ of the small-signal model involving the resonator model, the resulting currents, $I_{t}$ and $-I_{t}$, are given by

\begin{align}
I_{t}=&\underbrace{g_{mp}V'_{yp}+\frac{V_{x}}{Z_{dc}}}_{I_{xp}}+\underbrace{g_{mn}V'_{yn}+\frac{V_{x}}{Z_{dc}}}_{I_{xn}}+\underbrace{sC_{z}V_{x}}_{I_{xcz}}\\
{\nonumber}&+\underbrace{\frac{2V_{x}}{R_{p}}+\frac{2V_{x}}{sL_{p}}+s2C_{p}V_{x}}_{I_{xRLC}}\\
-I_{t}=&\underbrace{g_{mp}V'_{xp}+\frac{V_{y}}{Z_{dc}}}_{I_{yp}}+\underbrace{g_{mn}V'_{xn}+\frac{V_{y}}{Z_{dc}}}_{I_{yn}}+\underbrace{sC_{z}V_{y}}_{I_{ycz}}\\
{\nonumber}&+\underbrace{\frac{2V_{y}}{R_{p}}+\frac{2V_{y}}{sL_{p}}+s2C_{p}V_{y}}_{I_{yRLC}}
\end{align}

{\noindent}where $I_{xRLC}$ and $I_{yRLC}$ flow through the parallel RLC models branched from nodes $X$ and $Y$, respectively. To obtain the impedance $V_{t}/I_{t}$ of the \textit{2-Port Oscillator} involving the resonator model, $I_{t}$ and $-I_{t}$ are subtracted as follows:

\begin{align}
2I_{t}=&\left(g_{mp}+g_{mn}\right)\left(V_{y}-V_{x}\right)\frac{sR_{dc}C_{dc}}{1+sR_{dc}C_{dc}}\\
{\nonumber}&+\frac{sC_{dc}2\left(V_{x}-V_{y}\right)}{1+sR_{dc}C_{dc}}\\
{\nonumber}&+\left(sC_{z}+\frac{2}{R_{p}}+\frac{2}{sL_{p}}+s2C_{p}\right)\left(V_{x}-V_{y}\right)
\end{align}

Applying $V_{t}=V_{x}-V_{y}$ and $s=j{\omega}$ to the above equation and rearranging the result for $V_{t}/I_{t}$, the impedance, which consists of a real part (or resistance $R'_{t}$) and an imaginary part (or reactance $X'_{n}$), can be obtained as follows: \footnote{The derivations for $R'_{t}$ and $X'_{n}$ are provided in Appendix A.4, which conducts the small-signal analysis based on \cite{Lee2026osc}.}

\begin{align}
\frac{V_{t}}{I_{t}}=R'_{t}+jX'_{n}
\end{align}

Since each part of $R'_{t}+jX'_{n}$ is expressed as a complicated equation, unlike $L_{n}$ and $C_{n}$ of Equations (6) and (7) for Fig. \ref{fig. 3}, finding the inductive and capacitive components from $X'_{n}$ to obtain a resonant frequency can be a challenging task. As an alternative method, the resonant frequency can be derived by finding ${\omega}$ from $X'_{n}=0$. In general, the impedance of a series (or parallel) RLC circuit has a resonant frequency at a point where the reactance of its impedance becomes zero \cite{Lee2026osc}. Therefore, by calculating ${\omega}$ (or $2{\pi}f$) from $X'_{n}=0$ using Equation (29), the resonant frequency is derived as

\begin{align}
f&=\frac{1}{2{\pi}}\sqrt{\frac{G_{m}}{R_{dc}C_{dc}C_{xy}}+\frac{2}{L_{p}C_{xy}}+\frac{2}{R_{dc}C_{dc}R_{p}C_{xy}}}\\
&=\frac{1}{2{\pi}}\sqrt{\frac{1}{R_{dc}C_{dc}R_{Gm}C_{xy}}+\frac{2}{L_{p}C_{xy}}+\frac{2}{R_{dc}C_{dc}R_{p}C_{xy}}}
\end{align}

{\noindent}where $G_{m}=g_{mp}+g_{mn}$, $C_{xy}=C_{z}+2C_{p}$, and $G_{m}=1/R_{Gm}$ (see Appendix A.4 for the detailed derivation). If the small-signal model shown in Fig. \ref{fig.8} does not involve the resonator model by assuming $R_{p}={\infty}$, $L_{p}={\infty}$, and $C_{p}=0$ (i.e., treating the resonator as an open circuit), the resonant frequency in Equations (30) and (31) is approximated as that in Equations (9) and (10).

Recall that the impedance transformation for mBVD-to-parallel RLC model is conducted by assuming $L_{p}=L_{m}$, $C_{p}=C_{m}$, and $R_{p}=k_{t}/R_{s}$ as shown in Fig. \ref{fig. 7}, the resonant frequency in Equation (30) can be rewritten using the mBVD model parameters as follows:

\begin{align}
f=\frac{1}{2{\pi}}\sqrt{\left(\frac{G_{m}}{R_{dc}C_{dc}}+\frac{2}{L_{m}}+\frac{2R_{s}}{R_{dc}C_{dc}k_{t}}\right)\frac{1}{C_{z}+2C_{m}}}
\end{align}

{\noindent}where $L_{m}$ is the motional inductance, $C_{m}$ is the motional capacitance, $R_{s}$ represents the series electrode resistance, and $k_{t}$ is the transformation coefficient and depends on the resonator design, as described in Fig. \ref{fig. 7}.

Therefore, compared to the resonant frequency of the \textit{2-Port Oscillator} without the resonator model (Fig. \ref{fig. 3}), the resonant frequency of the \textit{2-Port Oscillator} incorporating the resonator model can be updated using the parameters $L_{m}$ and $C_{m}$, as well as $V_{p}$ and $V_{n}$ for changing $G_{m}$. 
\subsection{\color{myblue}Loss Effect on a Resonant Frequency}
In the mBVD model of Fig. \ref{fig. 7}, the series electrode resistance $R_{s}$ can vary depending on the physical dimensions and materials of a resonator (e.g., finger width, length, spacing in an interdigital transducer). Also, $R_{s}$ can be affected in interfacing with CMOS circuitry by pads, wire bonding, routing, and substrate loss. Consequently, $R_{s}$ results in the loss of a resonator and affects not only the resonant frequency but also the operation of a resonator-CMOS circuitry.

In the impedance transformation for the mBVD-to-parallel RLC model (Fig. \ref{fig. 7}), $R_{p}$ is modeled to be proportional to $1/R_{s}$ because a series resistance and a parallel resistance determine the impedance magnitude when a series RLC circuit and a parallel RLC circuit are respectively resonated. This means that, at the resonant frequencies of series and parallel RLC circuits, the series resistance forms a minimum peak magnitude while the parallel resistance forms a maximum peak magnitude, leading to the relationship $R_{p}{\propto}(1/R_{s})$ in the impedance transformation of this work. Therefore, the resonant frequency in Equation (30) is inversely proportional to $R_{p}$, whereas the resonant frequency in Equation (32) is proportional to $R_{s}$.

\begin{figure}[h!]
\centering\includegraphics[scale = 0.55]{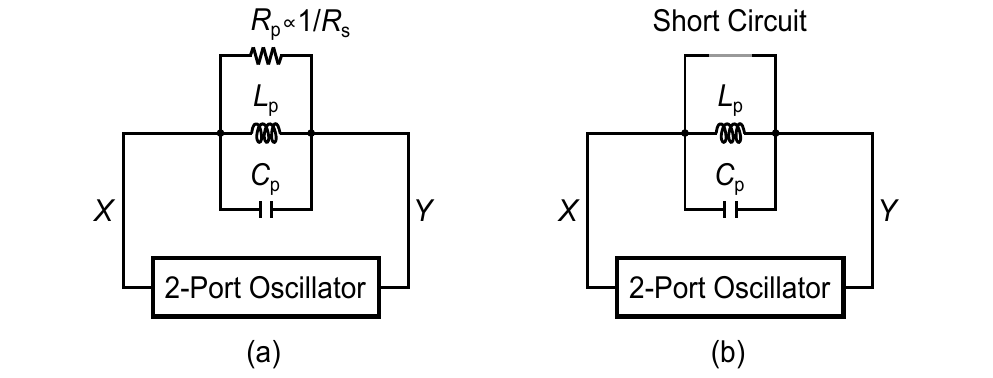}
\captionof{figure}{{\color{myred}\textbf{\textit{2-Port Oscillator} with a parallel RLC model.}} (a) If $R_{s}$ or $R_{p}$ has a finite value. (b) If $R_{s}$ approaches infinity or $R_{p}$ becomes a short circuit.}\label{fig.9}
\end{figure}

However, when designing a resonator and configuring an interface between the resonator and CMOS circuitry, $R_{s}$ may unintentionally increase due to device dimensions, pads, wire bonding, routing, and substrate loss. An increase in $R_{s}$ results in a decrease in $R_{p}$ according to $R_{p}{\propto}(1/R_{s})$. Considering the two extreme cases of $R_{s}$ or $R_{p}$ (Fig. \ref{fig.9}), if $R_{s}$ or $R_{p}$ has a finite value (Fig. \ref{fig.9}(a)), the \textit{2-Port Oscillator} generates a resonant frequency according to Equations (30) and (32). If $R_{s}$ continues to increase and approaches infinity, $R_{p}$ is eventually modeled as a short circuit, resulting in the vanishing of the oscillation (Fig. \ref{fig.9}(b)). Therefore, $R_{s}$ or $R_{p}$ must be properly controlled to minimize the loss and prevent the oscillation from dying out.

\subsection{\color{myblue}Coupled Oscillators Involving Resonator Models}
\begin{figure}[h!]
\centering\includegraphics[scale = 0.55]{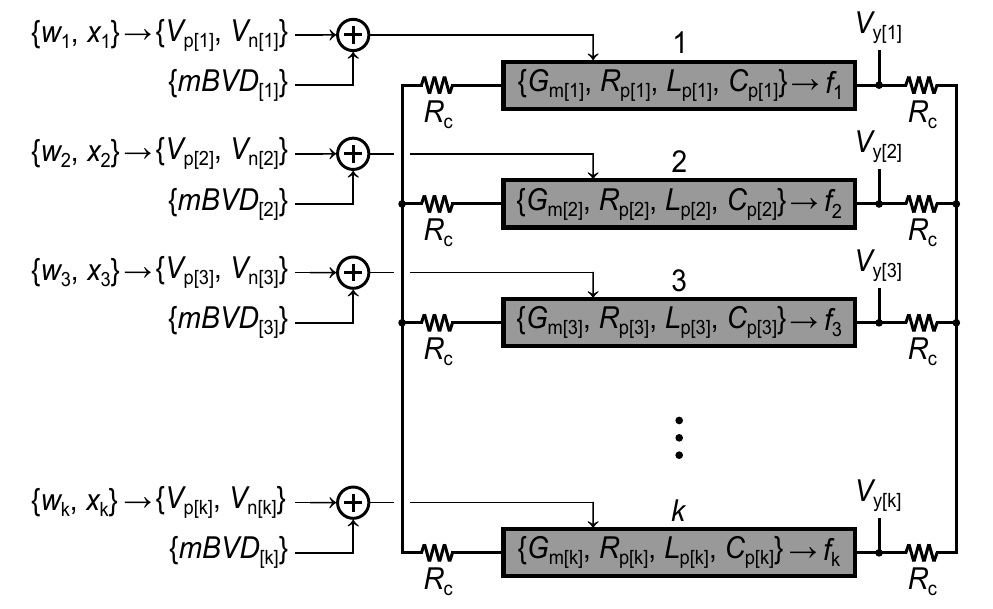}
\captionof{figure}{{\color{myred}\textbf{Coupled oscillators involving resonator models.}} Frequency synchronization model using \textit{2-Port Oscillator}s (gray boxes) with mBVD-to-parallel RLC models.}\label{fig.10}
\end{figure}

When the small-signal model described in Fig. \ref{fig.8} is expanded using an all-to-all connection scheme by a factor of $k$, which is the number of coupled oscillators, a frequency synchronization model can be implemented as shown in Fig. \ref{fig.10}. Compared to the model shown in Fig. \ref{fig. 4}, this frequency synchronization model includes the mBVD model parameters $\{mBVD_{[\alpha]}\}_{\alpha=1}^{k}$ (Fig. \ref{fig.10}). Each gray box represents the \textit{2-Port Oscillator}, which consists of the parameters $\{G_{m[\alpha]},R_{p[\alpha]},L_{p[\alpha]},C_{p[\alpha]}\}_{\alpha=1}^{k}$ for an individual resonant frequency $\{f_{\alpha}\}_{\alpha=1}^{k}$. The parameters $\{R_{p[\alpha]},L_{p[\alpha]},C_{p[\alpha]}\}_{\alpha=1}^{k}$ are obtained from $\{mBVD_{[\alpha]}\}_{\alpha=1}^{k}$ according to the mBVD-to-parallel RLC transformation performed in Fig. \ref{fig. 7}.

{\noindent}Note that $R_{p[\alpha]}$, $L_{p[\alpha]}$, and $C_{p[\alpha]}$ correspond to $R_{p}$, $L_{p}$, and $C_{p}$ shown in Fig. \ref{fig.8}, respectively. As with the model shown in Fig. \ref{fig. 4}, the datasets $\{w_{\alpha},x_{\alpha}\}_{\alpha=1}^{k}$ should be mapped to the voltage signals $\{V_{p[\alpha]},V_{n[\alpha]}\}_{\alpha=1}^{k}$ so that each multiplication $w_{\alpha}x_{\alpha}$ corresponds appropriately to a resonant frequency $f_{\alpha}$. $V_{p[\alpha]}$ and $V_{n[\alpha]}$ correspond to $V_{p}$ and $V_{n}$ shown in Fig. \ref{fig. 2}(a), respectively. In each \textit{2-Port Oscillator}, $V_{p[\alpha]}$ and $V_{n[\alpha]}$ control $g_{mp[\alpha]}$ and $g_{mn[\alpha]}$ of the upper and lower cross-coupled pairs, respectively, resulting in $\{G_{m[\alpha]}=g_{mp[\alpha]}+g_{mn[\alpha]}\}_{\alpha=1}^{k}$. $R_{c}$ controls the coupling strength between the coupled oscillators, as with the model shown in Fig. \ref{fig. 4}.

To simplify the analysis in Fig. \ref{fig.10}, the parameters of an each oscillator are assumed that $\{R_{p[\alpha]}\}_{\alpha=1}^{k}=R_{p}$, $\{L_{p[\alpha]}\}_{\alpha=1}^{k}=L_{p}$, and $\{C_{p[\alpha]}\}_{\alpha=1}^{k}=C_{p}$ so that all the oscillators are driven using the same parallel RLC model. Then, for finding the resonant frequency of the coupled oscillators shown in Fig. \ref{fig.10}, assuming that $R_{c}$ is low enough and applying a test voltage $V_{t}$ across the coupled oscillators, the resulting small-signal currents, $I_{t}$ and $-I_{t}$, flowing through the coupled oscillators are scaled by a factor of $k$ using Equations (26) and (27) as follows:

\begin{align}
I_{t}&=k\left(I_{xp}+I_{xn}+I_{xcz}+I_{xRLC}\right)\\
-I_{t}&=k\left(I_{yp}+I_{yn}+I_{ycz}+I_{yRLC}\right)
\end{align}

{\noindent}where the currents constituting $I_{t}$ and $-I_{t}$ are the same as those of the small-signal model shown in Fig. \ref{fig.8}. $I_{t}$ and $-I_{t}$ flow through the left and right sides of the coupled oscillators in Fig. \ref{fig.10}, respectively. Then, using Equations (26) and (27), the above currents are rewritten as

\begin{align}
I_{t}=&\underbrace{\sum_{\alpha=1}^{k}g_{mp[\alpha]}V'_{yp}+\frac{kV_{x}}{Z_{dc}}}_{k{\cdot}I_{xp}}+\underbrace{\sum_{\alpha=1}^{k}g_{mn[\alpha]}V'_{yn}+\frac{kV_{x}}{Z_{dc}}}_{k{\cdot}I_{xn}}\\
{\nonumber}&+\underbrace{sC_{z}kV_{x}}_{k{\cdot}I_{xcz}}+\underbrace{\frac{2kV_{x}}{R_{p}}+\frac{2kV_{x}}{sL_{p}}+s2C_{p}kV_{x}}_{k{\cdot}I_{xRLC}}\\
-I_{t}=&\underbrace{\sum_{\alpha=1}^{k}g_{mp[\alpha]}V'_{xp}+\frac{kV_{y}}{Z_{dc}}}_{k{\cdot}I_{yp}}+\underbrace{\sum_{\alpha=1}^{k}g_{mn[\alpha]}V'_{xn}+\frac{kV_{y}}{Z_{dc}}}_{k{\cdot}I_{yn}}\\
{\nonumber}&+\underbrace{sC_{z}kV_{y}}_{k{\cdot}I_{ycz}}+\underbrace{\frac{2kV_{y}}{R_{p}}+\frac{2kV_{y}}{sL_{p}}+s2C_{p}kV_{y}}_{k{\cdot}I_{yRLC}}
\end{align}

{\noindent}where the currents flowing through $r_{op}$ and $r_{on}$ depicted in Fig. \ref{fig.8} are neglected. As with the calculations for $V_{t}/I_{t}$ ($=R'_{t}+jX'_{n}$) in Equation (29), subtracting the above two currents, applying $V_{t}=V_{x}-V_{y}$ to the subtracted result, rearranging the resulting terms for $V_{t}/I_{t}$, and approximating $V_{t}/I_{t}$, the impedance across the coupled oscillators can be expressed using a real part (or resistance $R'_{tc}$) and an imaginary part (or reactance $X'_{nc}$) as follows: \footnote{The complete calculations for $V_{t}/I_{t}$ ($=R'_{tc}+jX'_{nc}$) and the resonant frequency (i.e., $f_{syn}$) are provided in Appendix A.5.}

\begin{align}
\frac{V_{t}}{I_{t}}&{\,\approx\,}\frac{2R_{p}L_{p}\left(-{\omega}^2R_{dc}C_{dc}+j{\omega}\right)}{\begin{array}{c}
\sum_{\alpha=1}^{k}G_{m[\alpha]}{\omega}^2R_{dc}C_{dc}R_{p}L_{p}\\
+j\left[\begin{array}{c}
-{\omega}^3R_{dc}C_{dc}R_{p}L_{p}\left(C_{z}+2C_{p}\right)k\\
+{\omega}2kR_{dc}C_{dc}R_{p}+{\omega}2kL_{p}
\end{array}\right]
\end{array}}\\
&=R'_{tc}+jX'_{nc}
\end{align}

{\noindent}where $G_{m[\alpha]}=g_{mp[\alpha]}+g_{mn[\alpha]}$. As with the derivation of the resonant frequency conducted in Section $\mathrm{III}.B$ (i.e., $X'_{n}=0$), the resonant frequency of the coupled oscillators can be derived by finding ${\omega}$ (or $2{\pi}f$) where the reactance $X'_{nc}$ of $V_{t}/I_{t}$ becomes zero. The following equation represents $X'_{nc}=0$ using the numerator of $X'_{nc}$.

\begin{align}
\underbrace{\left[\begin{array}{c}
\sum_{\alpha=1}^{k}G_{m[\alpha]}{\omega}^3R_{dc}C_{dc}R_{p}L_{p}\\
-{\omega}^5R_{dc}^2C_{dc}^2R_{p}L_{p}\left(C_{z}+2C_{p}\right)k\\
+{\omega}^3{2}kR_{dc}^2C_{dc}^2R_{p}+{\omega}^3{2}kR_{dc}C_{dc}L_{p}
\end{array}\right]}_{\text{Numerator of}{\,}X'_{nc}}=0
\end{align}

{\noindent}Therefore, from Equation (39), the resonant frequency corresponding to the synchronized frequency $f_{syn}$ of the coupled oscillators is obtained as follows:

\begin{align}
f_{syn}&=\frac{1}{2{\pi}}\sqrt{\frac{\sum_{\alpha=1}^{k}G_{m[\alpha]}}{R_{dc}C_{dc}C_{xy}k}+\underbrace{\frac{2}{L_{p}C_{xy}}+\frac{2}{R_{dc}C_{dc}R_{p}C_{xy}}}_{W}}\\
&=\frac{1}{2{\pi}}\sqrt{\sum_{\alpha=1}^{k}\frac{1}{R_{Gm[\alpha]}C_{xy}kR_{dc}C_{dc}}+W}
\end{align}

{\noindent}where $C_{xy}=C_{z}+2C_{p}$ and $G_{m[\alpha]}=1/R_{Gm[\alpha]}$. The above $f_{syn}$ expression is derived by assuming that all the oscillators in Fig. \ref{fig.10} are driven using the same parallel RLC model.

However, if all the coupled oscillators are driven employing individually controllable $C_{z}$ and parallel RLC parameters $\{R_{p[\alpha]},L_{p[\alpha]},C_{p[\alpha]}\}_{\alpha=1}^{k}$ shown in Fig. \ref{fig.10}, the effective impedances of $kC_{z}$, $R_{p}/k$, $L_{p}/k$, and $kC_{p}$ in Equations (35) and (36) must be modified as follows:

\begin{align}
kC_{z}&{\,\rightarrow\,}C'_{z}=\sum_{\alpha=1}^{k}C_{z[\alpha]}\\
\frac{R_{p}}{k}&{\,\rightarrow\,}R'_{p}=R_{p[1]}||R_{p[2]}{\cdots}||R_{p[k-1]}||R_{p[k]}\\
\frac{L_{p}}{k}&{\,\rightarrow\,}L'_{p}=L_{p[1]}||L_{p[2]}{\cdots}||L_{p[k-1]}||L_{p[k]}\\
kC_{p}&{\,\rightarrow\,}C'_{p}=\sum_{\alpha=1}^{k}C_{p[\alpha]}
\end{align}

{\noindent}where $k$ is the number of coupled oscillators, $C'_{z}$ is the effective load capacitance at the left or right side of the coupled oscillators (Fig. \ref{fig.10}), $R'_{p}$ is the effective parallel resistance for resonator models, $L'_{p}$ is the effective parallel inductance for resonator models, and $C'_{p}$ is the effective parallel capacitance for resonator models. Accordingly, the small-signal currents in Equations (35) and (36) are updated as

\begin{align}
I_{t}&=\underbrace{\sum_{\alpha=1}^{k}g_{mp[\alpha]}V'_{yp}+\frac{kV_{x}}{Z_{dc}}}_{k{\cdot}I_{xp}}+\underbrace{\sum_{\alpha=1}^{k}g_{mn[\alpha]}V'_{yn}+\frac{kV_{x}}{Z_{dc}}}_{k{\cdot}I_{xn}}\\
{\nonumber}&+\underbrace{sC'_{z}V_{x}}_{I'_{xcz}}+\underbrace{\frac{2V_{x}}{R'_{p}}+\frac{2V_{x}}{sL'_{p}}+s2C'_{p}V_{x}}_{I'_{xRLC}}\\
-I_{t}&=\underbrace{\sum_{\alpha=1}^{k}g_{mp[\alpha]}V'_{xp}+\frac{kV_{y}}{Z_{dc}}}_{k{\cdot}I_{yp}}+\underbrace{\sum_{\alpha=1}^{k}g_{mn[\alpha]}V'_{xn}+\frac{kV_{y}}{Z_{dc}}}_{k{\cdot}I_{yn}}\\
{\nonumber}&+\underbrace{sC'_{z}V_{y}}_{I'_{ycz}}+\underbrace{\frac{2V_{y}}{R'_{p}}+\frac{2V_{y}}{sL'_{p}}+s2C'_{p}V_{y}}_{I'_{yRLC}}
\end{align}

{\noindent}where $I'_{xcz}$ and $I'_{xRLC}$ are the currents flowing through $C'_{z}$ and the half-parallel RLC model, respectively, located on the left side of the coupled oscillators, while $I'_{ycz}$ and $I'_{yRLC}$ represent the currents flowing through the components on the right side of the coupled oscillators.

As conducted in the calculations for $V_{t}/I_{t}$ ($=R'_{tc}+jX'_{nc}$) and $f_{syn}$, the same calculations are performed from the updated small-signal currents in Equations (46) and (47) to obtain the updated impedance of $V_{t}/I_{t}$ ($=R''_{tc}+jX''_{nc}$), accordingly, the resonant frequency corresponding to the synchronized frequency $f_{syn}$ of the coupled oscillators in Fig. \ref{fig.10} is obtained as \footnote{By using the updated small-signal currents in Equations (46) and (47), the complete calculations for $V_{t}/I_{t}$ ($=R''_{tc}+jX''_{nc}$), $X''_{nc}=0$, and $f_{syn}$ are provided in Appendix A.5.}

\begin{align}
f_{syn}&=\frac{1}{2{\pi}}\sqrt{\frac{\sum_{\alpha=1}^{k}G_{m[\alpha]}}{R_{dc}C_{dc}C'_{xy}}+\underbrace{\frac{2}{L'_{p}C'_{xy}}+\frac{2}{R_{dc}C_{dc}R'_{p}C'_{xy}}}_{W'}}\\
&=\frac{1}{2{\pi}}\sqrt{\sum_{\alpha=1}^{k}\frac{1}{R_{Gm[\alpha]}C'_{xy}R_{dc}C_{dc}}+W'}
\end{align}

{\noindent}where $C'_{xy}=C'_{z}+2C'_{p}$ and $G_{m[\alpha]}=1/R_{Gm[\alpha]}$. Therefore, as shown in Fig. \ref{fig.10}, as $R_{c}$ is sufficiently high, the oscillators work with independent frequencies $\{f_{\alpha}\}_{\alpha=1}^{k}$ according to $\{C_{z[\alpha]}\}_{\alpha=1}^{k}$ and $\{G_{m[\alpha]},R_{p[\alpha]},L_{p[\alpha]},C_{p[\alpha]}\}_{\alpha=1}^{k}$. Then, as $R_{c}$ is reduced sufficiently, all the oscillators are synchronized with each other, and their independent frequencies $\{f_{\alpha}\}_{\alpha=1}^{k}$ are updated to $f_{syn}$ as defined in Equation (48).

The signal flow of the coupled oscillators in the synchronization mode is described in Equation (50). Compared to the signal flow that does not involve resonator models described in Equation (25), the function $\mathcal{T}_{c}(\cdot)$ is adopted for the mBVD-to-parallel RLC transformation. Then, the function $\mathcal{F}_{c}(\cdot)$ returns $f_{syn}$ using $\{G_{m[\alpha]},R_{p[\alpha]},L_{p[\alpha]},C_{p[\alpha]}\}_{\alpha=1}^{k}$, as defined in Equation (48).

\begin{align}
\begin{gathered}
\left\{w_{\alpha},x_{\alpha}\right\}_{\alpha=1}^{k}\\
\big\downarrow{\,\mathcal{N}_{c}(\cdot)}\\
\left\{V_{p[\alpha]},V_{n[\alpha]}\right\}_{\alpha=1}^{k}\\
\big\downarrow{\,\mathcal{M}_{c}(\cdot)}\\
\left\{G_{m[\alpha]}\right\}_{\alpha=1}^{k}\\
\big\downarrow{\,\mathcal{F}_{c}(\cdot)}\\
\boldsymbol{f_{syn}}\\
\big\uparrow{\,\mathcal{F}_{c}(\cdot)}\\
\left\{R_{p[\alpha]},L_{p[\alpha]},C_{p[\alpha]}\right\}_{\alpha=1}^{k}\\
\big\uparrow{\,\mathcal{T}_{c}(\cdot)}\\
\left\{mBVD_{[\alpha]}\right\}_{\alpha=1}^{k}
\end{gathered}
\end{align}

\begin{figure}[h!]
\centering\includegraphics[scale = 0.55]{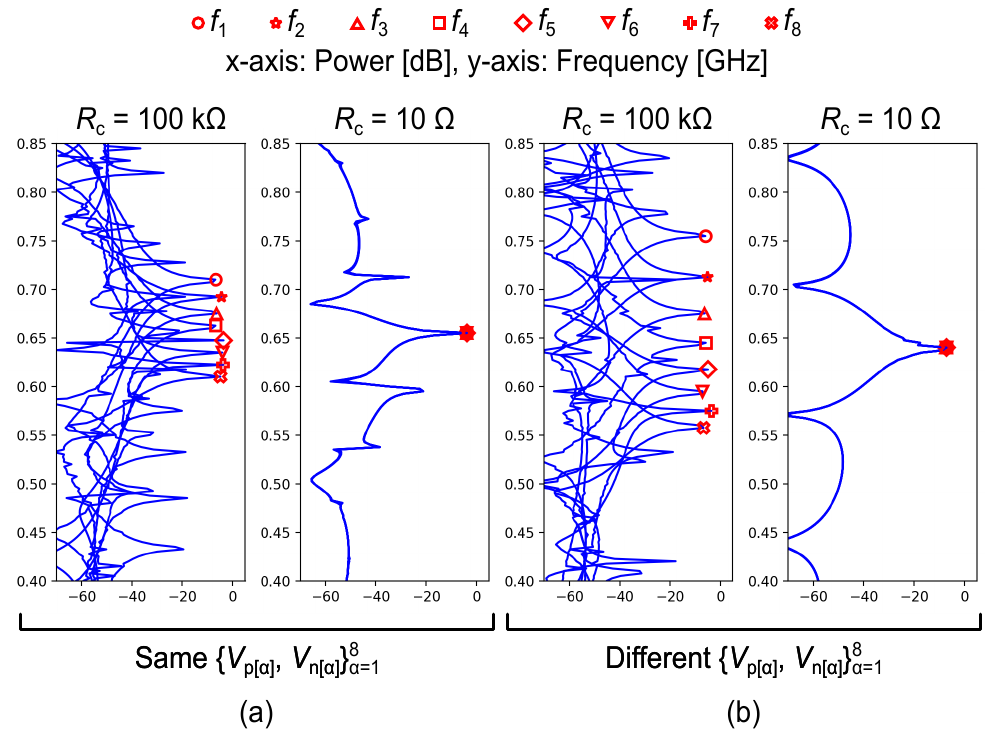}
\captionof{figure}{{\color{myred}\textbf{Frequency synchronization with parallel RLC models.}} Frequency variations of the respective oscillators in $k=8$ when the coupling resistor $R_{c}$ is updated from $100{\,}\mathrm{k}{\Omega}$ to $10{\,}{\Omega}$ according to (a) the same $\{V_{p[\alpha]},V_{n[\alpha]}\}_{\alpha=1}^{8}$ values and (b) the different $\{V_{p[\alpha]},V_{n[\alpha]}\}_{\alpha=1}^{8}$ values. The horizontal axis indicates $\mathrm{Power{\,}[dB]}$, and the vertical axis indicates $\mathrm{Frequency{\,}[GHz]}$.}\label{fig.11}
\end{figure}

Fig. \ref{fig.11} presents the frequency variations of the coupled oscillators shown in Fig. \ref{fig.10} according to the coupling resistor $R_{c}$. In setting the coupling mode, the value of $R_{c}$ is set to $100{\,}\mathrm{k}{\Omega}$ for the independent mode and set to $10{\,}{\Omega}$ for the synchronization mode. Using $k=8$, the frequencies $\{f_{\alpha}\}_{\alpha=1}^{k}$ of the coupled oscillators are obtained by the FFT of $\{V_{y[\alpha]}\}_{\alpha=1}^{k}$ shown in Fig. \ref{fig.10}. $f_{1}$ through $f_{8}$ are denoted using eight red symbols (Fig. \ref{fig.11}). The circuit implementation used to obtain Fig. \ref{fig.11} is described in Appendix A.3.

Fig. \ref{fig.11}(a) shows the frequency synchronization of the eight coupled oscillators when $R_{c}$ is updated from $100{\,}\mathrm{k}{\Omega}$ to $10{\,}{\Omega}$ while using the same $\{V_{p[\alpha]},V_{n[\alpha]}\}_{\alpha=1}^{k}$ values where $\{V_{p[\alpha]}\}_{\alpha=1}^{k}=0.1{\,}\mathrm{V}$ and $\{V_{n[\alpha]}\}_{\alpha=1}^{k}=1.1{\,}\mathrm{V}$ in each oscillator. Fig. \ref{fig.11}(b) shows the frequency synchronization when $R_{c}$ is updated while using the different $\{V_{p[\alpha]},V_{n[\alpha]}\}_{\alpha=1}^{k}$ values, which are the same as those of the setup in Appendix A.3. For both cases in Fig. \ref{fig.11}, the parameters of the parallel RLC models are used as follows: $\{R_{p[\alpha]}\}_{\alpha=1}^{8}=100{\,}\mathrm{k}{\Omega}$, $\{L_{p[\alpha]}\}_{\alpha=1}^{8}=100{\,}\mathrm{nF}$, and $\{C_{p[\alpha]}\}_{\alpha=1}^{8}=500\text{--}850{\,}\mathrm{fF}$ with a step of $50{\,}\mathrm{fF}$. Therefore, as the $R_{c}$ value is updated from $100{\,}\mathrm{k}{\Omega}$ to $10{\,}{\Omega}$, all the oscillators are synchronized with each other in both frequency and phase, as shown in Fig. \ref{fig.11}. 
\section{\color{myblue}\Large{P}\large{ARTIAL} \Large{S}\large{YNCHRONIZATION}}
In the frequency synchronization models described in Figs. \ref{fig. 4} and \ref{fig.10}, the coupling strength is controlled by applying the same value of $R_{c}$ to all oscillators. For example, in Fig. \ref{fig.11}, the $R_{c}$ values of the eight oscillators are set to $100{\,}\mathrm{k}{\Omega}$ for the independent mode and set to $10{\,}{\Omega}$ for the synchronization mode. Therefore, all the oscillators in Fig. \ref{fig.11} are simultaneously synchronized.

\begin{figure}[h!]
\centering\includegraphics[scale = 0.55]{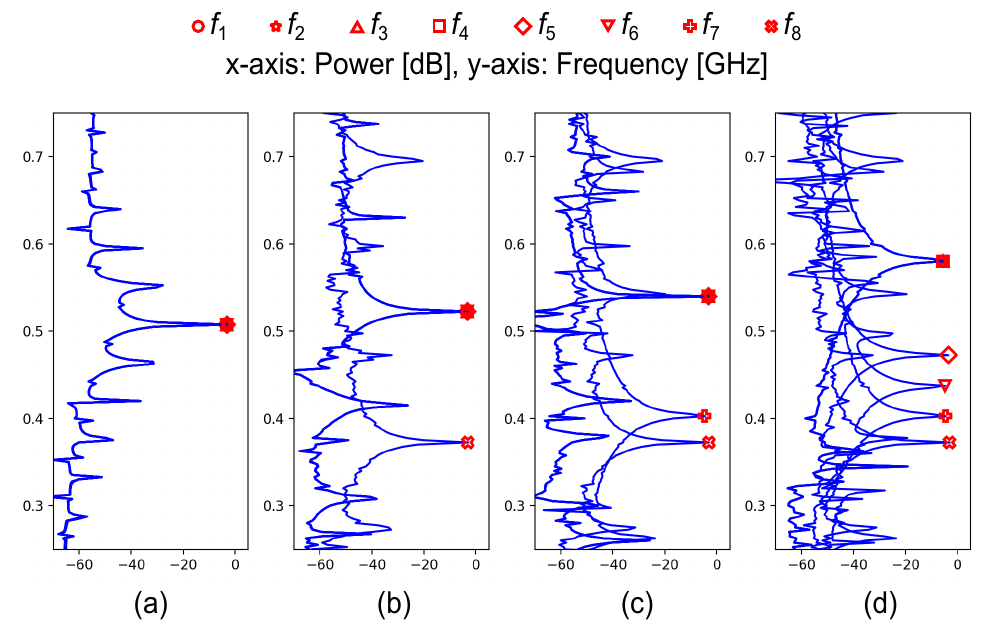}
\captionof{figure}{{\color{myred}\textbf{Partial synchronization without parallel RLC models.}} Frequencies of the eight oscillators when (a) $\{R_{c[\alpha]}\}_{\alpha=1}^{8}=10{\,}{\Omega}$; (b) $\{R_{c[\alpha]}\}_{\alpha=1}^{7}=10{\,}{\Omega}$ and $R_{c[8]}=100{\,}\mathrm{k}{\Omega}$; (c) $\{R_{c[\alpha]}\}_{\alpha=1}^{6}=10{\,}{\Omega}$ and $\{R_{c[\alpha]}\}_{\alpha=7}^{8}=100{\,}\mathrm{k}{\Omega}$; and (d) $\{R_{c[\alpha]}\}_{\alpha=1}^{4}=10{\,}{\Omega}$ and $\{R_{c[\alpha]}\}_{\alpha=5}^{8}=100{\,}\mathrm{k}{\Omega}$.}\label{fig.12}
\end{figure}

However, if the coupling strength of each oscillator can be independently controlled using a coupling resistor, the synchronization can be partially achieved using a selected subset of oscillators. Therefore, in this section, a dedicated coupling resistor $R_{c[\alpha]}$ is assigned to each oscillator for the partial synchronization ($\alpha$ is the index number for the $k$ oscillators). Using $k=8$, the coupling strengths of the eight oscillators are individually controlled by tuning $R_{c[1]}$ through $R_{c[8]}$, expressed as $\{R_{c[\alpha]}\}_{\alpha=1}^{8}$. That is, in Figs. \ref{fig. 4} and \ref{fig.10}, $R_{c[1]}$ controls the coupling of $f_{1}$, and $R_{c[k]}$ controls the coupling of $f_{k}$.

Using $k=8$, Fig. \ref{fig.12} shows the frequency variations of the eight oscillators according to $\{R_{c[\alpha]}\}_{\alpha=1}^{8}$ in the frequency synchronization model described in Fig. \ref{fig. 4}. When $\{R_{c[\alpha]}\}_{\alpha=1}^{8}=10{\,}{\Omega}$, all the frequencies $\{f_{\alpha}\}_{\alpha=1}^{8}$ are synchronized each other (Fig. \ref{fig.12}(a)), i.e., $\{f_{\alpha}\}_{\alpha=1}^{8}=f_{syn}$ as defined in Equation (23). As the coupling strength of the eighth oscillator weakens by changing $R_{c[8]}$ from $10{\,}{\Omega}$ to $100{\,}\mathrm{k}{\Omega}$, $f_{8}$ is decoupled from $f_{syn}$ while sustaining $\{f_{\alpha}\}_{\alpha=1}^{7}=f_{syn}$ (Fig. \ref{fig.12}(b)). As the coupling strengths of the seventh and eighth oscillators weaken by changing $\{R_{c[\alpha]}\}_{\alpha=7}^{8}$ from $10{\,}{\Omega}$ to $100{\,}\mathrm{k}{\Omega}$, $f_{7}$ and $f_{8}$ are decoupled from $f_{syn}$ while sustaining $\{f_{\alpha}\}_{\alpha=1}^{6}=f_{syn}$ (Fig. \ref{fig.12}(c)). Similarly, as the coupling strengths of the fifth through eighth oscillators weaken by changing $\{R_{c[\alpha]}\}_{\alpha=5}^{8}$ to $100{\,}\mathrm{k}{\Omega}$, the frequencies $f_{5}$ through $f_{8}$ are decoupled from $f_{syn}$ while the remaining half of the oscillators maintain synchronization (Fig. \ref{fig.12}(d)). Fig. \ref{fig.12} is obtained based on the circuit described in Appendix A.3, and the values of $\{V_{p[\alpha]},V_{n[\alpha]}\}_{\alpha=1}^{8}$ are the same as those of the setup for Fig. \ref{fig. 5}.

\begin{figure}[h!]
\centering\includegraphics[scale = 0.55]{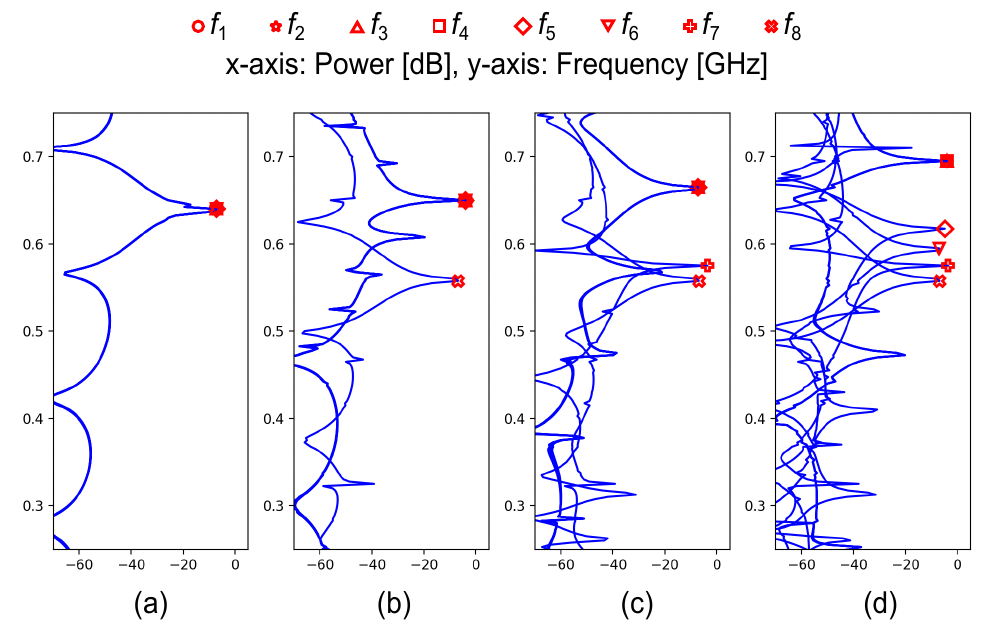}
\captionof{figure}{{\color{myred}\textbf{Partial synchronization with parallel RLC models.}} Frequencies of the eight oscillators when (a) $\{R_{c[\alpha]}\}_{\alpha=1}^{8}=10{\,}{\Omega}$; (b) $\{R_{c[\alpha]}\}_{\alpha=1}^{7}=10{\,}{\Omega}$ and $R_{c[8]}=100{\,}\mathrm{k}{\Omega}$; (c) $\{R_{c[\alpha]}\}_{\alpha=1}^{6}=10{\,}{\Omega}$ and $\{R_{c[\alpha]}\}_{\alpha=7}^{8}=100{\,}\mathrm{k}{\Omega}$; and (d) $\{R_{c[\alpha]}\}_{\alpha=1}^{4}=10{\,}{\Omega}$ and $\{R_{c[\alpha]}\}_{\alpha=5}^{8}=100{\,}\mathrm{k}{\Omega}$.}\label{fig.13}
\end{figure}

Fig. \ref{fig.13} shows the partial synchronization of the coupled oscillators involving resonator models described in Fig. \ref{fig.10}. The partial synchronization in Fig. \ref{fig.13} is conducted using the circuit described in Appendix A.3. The values of $\{V_{p[\alpha]},V_{n[\alpha]}\}_{\alpha=1}^{8}$ are identical to the setup for Fig. \ref{fig. 5}. The parameters of the parallel RLC models are $\{R_{p[\alpha]}\}_{\alpha=1}^{8}=100{\,}\mathrm{k}{\Omega}$, $\{L_{p[\alpha]}\}_{\alpha=1}^{8}=100{\,}\mathrm{nF}$, and $\{C_{p[\alpha]}\}_{\alpha=1}^{8}=500\text{--}850{\,}\mathrm{fF}$ with a step of $50{\,}\mathrm{fF}$. When all the coupling resistors are set to $10{\,}{\Omega}$ (i.e., $\{R_{c[\alpha]}\}_{\alpha=1}^{8}=10{\,}{\Omega}$), all the frequencies $\{f_{\alpha}\}_{\alpha=1}^{8}$ are synchronized to $f_{syn}$ as shown in Fig. \ref{fig.13}(a). Then, as $R_{c[5]}$ through $R_{c[8]}$ sequentially change from $10{\,}{\Omega}$ to $100{\,}\mathrm{k}{\Omega}$, $f_{5}$ through $f_{8}$ are sequentially decoupled from $f_{syn}$ as shown in Figs. \ref{fig.13}(b), (c), and (d).

In the results of Figs. \ref{fig.12} and \ref{fig.13}, the partial synchronization for a selected subset of oscillators is achieved by controlling the individual coupling resistors $\{R_{c[\alpha]}\}_{\alpha=1}^{8}$. Therefore, by isolating oscillators with functional defects from normal oscillators, the impact of the malfunctioning oscillators on the synchronization can be minimized. 
\section{\color{myblue}\Large{C}\large{ONCLUSION}}
This work provides a comprehensive theoretical circuit analysis of the frequency synchronization circuit model for analog computing in the frequency domain. Section $\mathrm{II}$ provides the analysis of a single oscillator and the analysis of the coupled oscillators that achieve synchronization without resonator models. Section $\mathrm{III}$ provides the impedance transformation for the mBVD-to-parallel RLC model and the analysis of the coupled oscillators involving resonator models. Section $\mathrm{IV}$ covers the partial synchronization of the coupled oscillators without (or with) the resonator models by individually controlling the coupling resistors.

The analysis and verification of the frequency synchronization circuit model are conducted using Cadence Virtuoso Studio (IC23.1-64b.43), and the single oscillator is designed using the complementary cross-coupled transconductance pairs and RC high-pass filters based on a $130{\,}\mathrm{nm}$ CMOS process design kit (PDK).

\section{\color{myblue}\Large{A}\large{CKNOWLEDGMENT}}
The authors acknowledge the financial support from the Defense Advanced Research Projects Agency (DARPA) via the NanoWatt Platforms for Sensing, Analysis and Computation (NaPSAC) Program under Grant N660012424004. 
\pretocmd{\thebibliography}{\color{mygreen}}{}{}
\balance

\balance

\newpage
\onecolumn
\section{\color{myblue}\Large{A}\large{PPENDIX}}
\renewcommand{\thefigure}{S\arabic{figure}}
\setcounter{figure}{0}

\noindent\textbf{\color{myblue}\textit{A}.1.\space{\,}Details of the Small-Signal Analysis for the \textit{2-Port Oscillator}}\\
{\indent}From Fig. \ref{fig. 3}, using $V'_{xp}=V'_{xn}=sR_{dc}C_{dc}V_{x}/(1+sR_{dc}C_{dc})$, $V'_{yp}=V'_{yn}=sR_{dc}C_{dc}V_{y}/(1+sR_{dc}C_{dc})$, and $Z_{dc}=(1+sR_{dc}C_{dc})/sC_{dc}$, the KCL results at nodes $X$ and $Y$ are given by

\begin{align}
I_{t}&=g_{mp}V'_{yp}+\frac{V_{x}}{Z_{dc}}+g_{mn}V'_{yn}+\frac{V_{x}}{Z_{dc}}+sC_{z}V_{x}\\
&=g_{mp}V_{y}\frac{sR_{dc}C_{dc}}{1+sR_{dc}C_{dc}}+\frac{sC_{dc}V_{x}}{1+sR_{dc}C_{dc}}+g_{mn}V_{y}\frac{sR_{dc}C_{dc}}{1+sR_{dc}C_{dc}}+\frac{sC_{dc}V_{x}}{1+sR_{dc}C_{dc}}+sC_{z}V_{x}\\
&=\left(g_{mp}+g_{mn}\right)V_{y}\frac{sR_{dc}C_{dc}}{1+sR_{dc}C_{dc}}+\frac{sC_{dc}2V_{x}}{1+sR_{dc}C_{dc}}+sC_{z}V_{x}\\
-I_{t}&=g_{mp}V'_{xp}+\frac{V_{y}}{Z_{dc}}+g_{mn}V'_{xn}+\frac{V_{y}}{Z_{dc}}+sC_{z}V_{y}\\
&=g_{mp}V_{x}\frac{sR_{dc}C_{dc}}{1+sR_{dc}C_{dc}}+\frac{sC_{dc}V_{y}}{1+sR_{dc}C_{dc}}+g_{mn}V_{x}\frac{sR_{dc}C_{dc}}{1+sR_{dc}C_{dc}}+\frac{sC_{dc}V_{y}}{1+sR_{dc}C_{dc}}+sC_{z}V_{y}\\
&=\left(g_{mp}+g_{mn}\right)V_{x}\frac{sR_{dc}C_{dc}}{1+sR_{dc}C_{dc}}+\frac{sC_{dc}2V_{y}}{1+sR_{dc}C_{dc}}+sC_{z}V_{y}
\end{align}

{\noindent}Note that $V'_{xp}$, $V'_{xn}$, $V'_{yp}$, and $V'_{yn}$ are obtained using the voltage division of $R_{dc}$ and $C_{dc}$, $Z_{dc}$ is the series impedance of $R_{dc}$ and $C_{dc}$. From the above equations, the subtraction of $I_{t}$ and $-I_{t}$ is

\begin{align}
2I_{t}=\left(g_{mp}+g_{mn}\right)\left(V_{y}-V_{x}\right)\frac{sR_{dc}C_{dc}}{1+sR_{dc}C_{dc}}+\frac{sC_{dc}2\left(V_{x}-V_{y}\right)}{1+sR_{dc}C_{dc}}+sC_{z}\left(V_{x}-V_{y}\right)
\end{align}

{\noindent}Using $V_{t}=V_{x}-V_{y}$, the above equation is reduced to

\begin{align}
2I_{t}=-V_{t}\left(g_{mp}+g_{mn}\right)\frac{sR_{dc}C_{dc}}{1+sR_{dc}C_{dc}}+\frac{sC_{dc}2V_{t}}{1+sR_{dc}C_{dc}}+sC_{z}V_{t}
\end{align}

{\noindent}The impedance $V_{t}/I_{t}$ is therefore expressed as follows:

\begin{align}
\frac{V_{t}}{I_{t}}&=\frac{2\left(1+sR_{dc}C_{dc}\right)}{-\left(g_{mp}+g_{mn}\right)sR_{dc}C_{dc}+s2C_{dc}+sC_{z}\left(1+sR_{dc}C_{dc}\right)}\\
&=\frac{2\left(1+j{\omega}R_{dc}C_{dc}\right)}{-{\omega}^2R_{dc}C_{dc}C_{z}+j{\omega}\left[-\left(g_{mp}+g_{mn}\right)R_{dc}C_{dc}+2C_{dc}+C_{z}\right]}
\end{align}

{\noindent}From the above expression, multiplying the numerator and denominator by the complex conjugate of the denominator yields the following:

\begin{align}
\frac{2\left(1+j{\omega}R_{dc}C_{dc}\right)\left\{-{\omega}^2R_{dc}C_{dc}C_{z}+j{\omega}\left[\left(g_{mp}+g_{mn}\right)R_{dc}C_{dc}-2C_{dc}-C_{z}\right]\right\}}{{\omega}^4\left(R_{dc}C_{dc}C_{z}\right)^2+{\omega}^2\left[-\left(g_{mp}+g_{mn}\right)R_{dc}C_{dc}+2C_{dc}+C_{z}\right]^2}
\end{align}

{\noindent}By designing the oscillator to be $\left(g_{mp}+g_{mn}\right)R_{dc}C_{dc}{\gg}\left(2C_{dc}+C_{z}\right)$, the above equation for $V_{t}/I_{t}$ can be approximated as

\begin{align}
\frac{2\left(1+j{\omega}R_{dc}C_{dc}\right)\left[-{\omega}^2R_{dc}C_{dc}C_{z}+j{\omega}\left(g_{mp}+g_{mn}\right)R_{dc}C_{dc}\right]}{{\omega}^4\left(R_{dc}C_{dc}C_{z}\right)^2+{\omega}^2\left[\left(g_{mp}+g_{mn}\right)R_{dc}C_{dc}\right]^2}
\end{align}

{\noindent}Then, $V_{t}/I_{t}$ can be expressed with the resistance $R_{t}$ and reactance $X_{t}$ as follows:

\begin{align}
\frac{V_{t}}{I_{t}}=\underbrace{\frac{-2{\omega}^2R_{dc}C_{dc}C_{z}-2{\omega}^2\left(g_{mp}+g_{mn}\right)R_{dc}^2C_{dc}^2}{{\omega}^4\left(R_{dc}C_{dc}C_{z}\right)^2+{\omega}^2\left[\left(g_{mp}+g_{mn}\right)R_{dc}C_{dc}\right]^2}}_{R_{t}}+j\underbrace{\frac{-2{\omega}^3R_{dc}^2C_{dc}^2C_{z}+2{\omega}\left(g_{mp}+g_{mn}\right)R_{dc}C_{dc}}{{\omega}^4\left(R_{dc}C_{dc}C_{z}\right)^2+{\omega}^2\left[\left(g_{mp}+g_{mn}\right)R_{dc}C_{dc}\right]^2}}_{X_{t}}
\end{align}

{\noindent}By choosing $1/({\omega}C_{z})$ to be larger than $1/(g_{mp}+g_{mn})$, the denominators of $R_{t}$ and $X_{t}$ can be approximated as follows:

\begin{align}
\frac{V_{t}}{I_{t}}&{\,\approx\,}\underbrace{\frac{-2{\omega}^2R_{dc}C_{dc}C_{z}-2{\omega}^2\left(g_{mp}+g_{mn}\right)R_{dc}^2C_{dc}^2}{{\omega}^2\left[\left(g_{mp}+g_{mn}\right)R_{dc}C_{dc}\right]^2}}_{R_{t}}+j\underbrace{\frac{-2{\omega}^3R_{dc}^2C_{dc}^2C_{z}+2{\omega}\left(g_{mp}+g_{mn}\right)R_{dc}C_{dc}}{{\omega}^2\left[\left(g_{mp}+g_{mn}\right)R_{dc}C_{dc}\right]^2}}_{X_{t}}\\
&=\frac{-2R_{dc}C_{dc}C_{z}-2\left(g_{mp}+g_{mn}\right)R_{dc}^2C_{dc}^2}{\left(g_{mp}+g_{mn}\right)^2R_{dc}^2C_{dc}^2}+j\left[\frac{-2{\omega}C_{z}}{\left(g_{mp}+g_{mn}\right)^2}+\frac{2}{{\omega}\left(g_{mp}+g_{mn}\right)R_{dc}C_{dc}}\right]\\
&{\,\approx\,}\frac{-2\left(g_{mp}+g_{mn}\right)R_{dc}^2C_{dc}^2}{\left(g_{mp}+g_{mn}\right)^2R_{dc}^2C_{dc}^2}+j\left[\frac{-2{\omega}C_{z}}{\left(g_{mp}+g_{mn}\right)^2}+\frac{2}{{\omega}\left(g_{mp}+g_{mn}\right)R_{dc}C_{dc}}\right]\\
&=\frac{-2}{g_{mp}+g_{mn}}+j{\omega}\frac{-2C_{z}}{\left(g_{mp}+g_{mn}\right)^2}+\frac{2}{-j{\omega}\left(g_{mp}+g_{mn}\right)R_{dc}C_{dc}}\\
&=R_{t}+j{\omega}L_{n}+\frac{1}{j{\omega}C_{n}}
\end{align}

As a result of the above equation, the impedance between nodes $X$ and $Y$ in the \textit{2-Port Oscillator} is derived as a series impedance consisting of a negative resistance $R_{t}$, a negative inductance $L_{n}$, and a negative capacitance $C_{n}$, where each element is given as follows:

\begin{align}
R_{t}&=\frac{-2}{g_{mp}+g_{mn}}\\
L_{n}&=\frac{-2C_{z}}{\left(g_{mp}+g_{mn}\right)^2}\\
C_{n}&=\frac{-\left(g_{mp}+g_{mn}\right)R_{dc}C_{dc}}{2}
\end{align} 
\noindent\textbf{\color{myblue}\textit{A}.2.\space{\,}Details of the Small-Signal Analysis for the Coupled Oscillators}\\
{\indent}Using $V'_{xp}=V'_{xn}=sR_{dc}C_{dc}V_{x}/(1+sR_{dc}C_{dc})$, $V'_{yp}=V'_{yn}=sR_{dc}C_{dc}V_{y}/(1+sR_{dc}C_{dc})$, and $Z_{dc}=(1+sR_{dc}C_{dc})/sC_{dc}$, $I_{t}$ and $-I_{t}$ in Equations (15) and (16) are rewritten as

\begin{align}
I_{t}&=\underbrace{\sum_{\alpha=1}^{k}g_{mp[\alpha]}V'_{yp}+\frac{kV_{x}}{Z_{dc}}}_{k{\cdot}I_{xp}}+\underbrace{\sum_{\alpha=1}^{k}g_{mn[\alpha]}V'_{yn}+\frac{kV_{x}}{Z_{dc}}}_{k{\cdot}I_{xn}}+\underbrace{sC_{z}kV_{x}}_{k{\cdot}I_{xcz}}\\
&=\sum_{\alpha=1}^{k}g_{mp[\alpha]}V_{y}\frac{sR_{dc}C_{dc}}{1+sR_{dc}C_{dc}}+\frac{sC_{dc}kV_{x}}{1+sR_{dc}C_{dc}}+\sum_{\alpha=1}^{k}g_{mn[\alpha]}V_{y}\frac{sR_{dc}C_{dc}}{1+sR_{dc}C_{dc}}+\frac{sC_{dc}kV_{x}}{1+sR_{dc}C_{dc}}+sC_{z}kV_{x}\\
&=\sum_{\alpha=1}^{k}G_{m[\alpha]}V_{y}\frac{sR_{dc}C_{dc}}{1+sR_{dc}C_{dc}}+\frac{sC_{dc}2kV_{x}}{1+sR_{dc}C_{dc}}+sC_{z}kV_{x}\\
-I_{t}&=\underbrace{\sum_{\alpha=1}^{k}g_{mp[\alpha]}V'_{xp}+\frac{kV_{y}}{Z_{dc}}}_{k{\cdot}I_{yp}}+\underbrace{\sum_{\alpha=1}^{k}g_{mn[\alpha]}V'_{xn}+\frac{kV_{y}}{Z_{dc}}}_{k{\cdot}I_{yn}}+\underbrace{sC_{z}kV_{y}}_{k{\cdot}I_{ycz}}\\
&=\sum_{\alpha=1}^{k}g_{mp[\alpha]}V_{x}\frac{sR_{dc}C_{dc}}{1+sR_{dc}C_{dc}}+\frac{sC_{dc}kV_{y}}{1+sR_{dc}C_{dc}}+\sum_{\alpha=1}^{k}g_{mn[\alpha]}V_{x}\frac{sR_{dc}C_{dc}}{1+sR_{dc}C_{dc}}+\frac{sC_{dc}kV_{y}}{1+sR_{dc}C_{dc}}+sC_{z}kV_{y}\\
&=\sum_{\alpha=1}^{k}G_{m[\alpha]}V_{x}\frac{sR_{dc}C_{dc}}{1+sR_{dc}C_{dc}}+\frac{sC_{dc}2kV_{y}}{1+sR_{dc}C_{dc}}+sC_{z}kV_{y}
\end{align}

{\noindent}where $G_{m[\alpha]}=g_{mp[\alpha]}+g_{mn[\alpha]}$, and $k$ is the number of coupled oscillators. From the above equations of $I_{t}$ and $-I_{t}$, the subtraction of the two currents is given by

\begin{align}
2I_{t}=\sum_{\alpha=1}^{k}G_{m[\alpha]}\left(V_{y}-V_{x}\right)\frac{sR_{dc}C_{dc}}{1+sR_{dc}C_{dc}}+\frac{sC_{dc}2k\left(V_{x}-V_{y}\right)}{1+sR_{dc}C_{dc}}+sC_{z}k\left(V_{x}-V_{y}\right)
\end{align}

{\noindent}Using $V_{t}=V_{x}-V_{y}$, the above equation is rewritten as

\begin{align}
2I_{t}=-\sum_{\alpha=1}^{k}G_{m[\alpha]}V_{t}\frac{sR_{dc}C_{dc}}{1+sR_{dc}C_{dc}}+\frac{sC_{dc}2kV_{t}}{1+sR_{dc}C_{dc}}+sC_{z}kV_{t}
\end{align}

{\noindent}Therefore, the impedance $V_{t}/I_{t}$ is obtained as follows:

\begin{align}
\frac{V_{t}}{I_{t}}&=\frac{2\left(1+sR_{dc}C_{dc}\right)}{-\sum_{\alpha=1}^{k}G_{m[\alpha]}sR_{dc}C_{dc}+sC_{dc}2k+sC_{z}k\left(1+sR_{dc}C_{dc}\right)}\\
&=\frac{2\left(1+j{\omega}R_{dc}C_{dc}\right)}{-{\omega}^2R_{dc}C_{dc}C_{z}k+j{\omega}\left(-\sum_{\alpha=1}^{k}G_{m[\alpha]}R_{dc}C_{dc}+C_{dc}2k+C_{z}k\right)}
\end{align}

{\noindent}Multiplying the numerator and denominator of the above $V_{t}/I_{t}$ by the complex conjugate of the denominator is given by

\begin{align}
\frac{2\left(1+j{\omega}R_{dc}C_{dc}\right)\left[-{\omega}^2R_{dc}C_{dc}C_{z}k+j{\omega}\left(\sum_{\alpha=1}^{k}G_{m[\alpha]}R_{dc}C_{dc}-C_{dc}2k-C_{z}k\right)\right]}{{\omega}^4\left(R_{dc}C_{dc}C_{z}k\right)^2+{\omega}^2\left(-\sum_{\alpha=1}^{k}G_{m[\alpha]}R_{dc}C_{dc}+C_{dc}2k+C_{z}k\right)^2}
\end{align}

{\noindent}By designing each oscillator in Fig. \ref{fig. 4} to be $G_{m[\alpha]}R_{dc}C_{dc}{\gg}\left(2C_{dc}+C_{z}\right)$, the above expression can be approximated as

\begin{align}
\frac{2\left(1+j{\omega}R_{dc}C_{dc}\right)\left(-{\omega}^2R_{dc}C_{dc}C_{z}k+j{\omega}\sum_{\alpha=1}^{k}G_{m[\alpha]}R_{dc}C_{dc}\right)}{{\omega}^4\left(R_{dc}C_{dc}C_{z}k\right)^2+{\omega}^2\left(\sum_{\alpha=1}^{k}G_{m[\alpha]}R_{dc}C_{dc}\right)^2}
\end{align}

{\noindent}Then, the above $V_{t}/I_{t}$ can be rewritten as the resistance $R_{tc}$ and reactance $X_{tc}$ as follows:

\begin{align}
\frac{V_{t}}{I_{t}}=\underbrace{\frac{-2{\omega}^2R_{dc}C_{dc}C_{z}k-2{\omega}^2\sum_{\alpha=1}^{k}G_{m[\alpha]}R_{dc}^2C_{dc}^2}{{\omega}^4\left(R_{dc}C_{dc}C_{z}k\right)^2+{\omega}^2\left(\sum_{\alpha=1}^{k}G_{m[\alpha]}R_{dc}C_{dc}\right)^2}}_{R_{tc}}+j\underbrace{\frac{-2{\omega}^3R_{dc}^2C_{dc}^2C_{z}k+2{\omega}\sum_{\alpha=1}^{k}G_{m[\alpha]}R_{dc}C_{dc}}{{\omega}^4\left(R_{dc}C_{dc}C_{z}k\right)^2+{\omega}^2\left(\sum_{\alpha=1}^{k}G_{m[\alpha]}R_{dc}C_{dc}\right)^2}}_{X_{tc}}
\end{align}

{\noindent}By choosing $1/({\omega}C_{z})$ to be larger than $1/G_{m[\alpha]}$, the denominators of $R_{tc}$ and $X_{tc}$ can be approximated as follows:

\begin{align}
\frac{V_{t}}{I_{t}}&{\,\approx\,}\underbrace{\frac{-2{\omega}^2R_{dc}C_{dc}C_{z}k-2{\omega}^2\sum_{\alpha=1}^{k}G_{m[\alpha]}R_{dc}^2C_{dc}^2}{{\omega}^2\left(\sum_{\alpha=1}^{k}G_{m[\alpha]}R_{dc}C_{dc}\right)^2}}_{R_{tc}}+j\underbrace{\frac{-2{\omega}^3R_{dc}^2C_{dc}^2C_{z}k+2{\omega}\sum_{\alpha=1}^{k}G_{m[\alpha]}R_{dc}C_{dc}}{{\omega}^2\left(\sum_{\alpha=1}^{k}G_{m[\alpha]}R_{dc}C_{dc}\right)^2}}_{X_{tc}}\\
&=\frac{-2R_{dc}C_{dc}C_{z}k-2\sum_{\alpha=1}^{k}G_{m[\alpha]}R_{dc}^2C_{dc}^2}{\left(\sum_{\alpha=1}^{k}G_{m[\alpha]}R_{dc}C_{dc}\right)^2}+j\left[\frac{-2{\omega}C_{z}k}{\left(\sum_{\alpha=1}^{k}G_{m[\alpha]}\right)^2}+\frac{2}{{\omega}\sum_{\alpha=1}^{k}G_{m[\alpha]}R_{dc}C_{dc}}\right]\\
&{\,\approx\,}\frac{-2\sum_{\alpha=1}^{k}G_{m[\alpha]}R_{dc}^2C_{dc}^2}{\left(\sum_{\alpha=1}^{k}G_{m[\alpha]}R_{dc}C_{dc}\right)^2}+j\left[\frac{-2{\omega}C_{z}k}{\left(\sum_{\alpha=1}^{k}G_{m[\alpha]}\right)^2}+\frac{2}{{\omega}\sum_{\alpha=1}^{k}G_{m[\alpha]}R_{dc}C_{dc}}\right]\\
&=\frac{-2}{\sum_{\alpha=1}^{k}G_{m[\alpha]}}+j{\omega}\frac{-2C_{z}k}{\left(\sum_{\alpha=1}^{k}G_{m[\alpha]}\right)^2}+\frac{2}{-j{\omega}\sum_{\alpha=1}^{k}G_{m[\alpha]}R_{dc}C_{dc}}\\
&=R_{tc}+j{\omega}L_{nc}+\frac{1}{j{\omega}C_{nc}}
\end{align}

\begin{figure}[h!]
\centering\includegraphics[scale = 0.55]{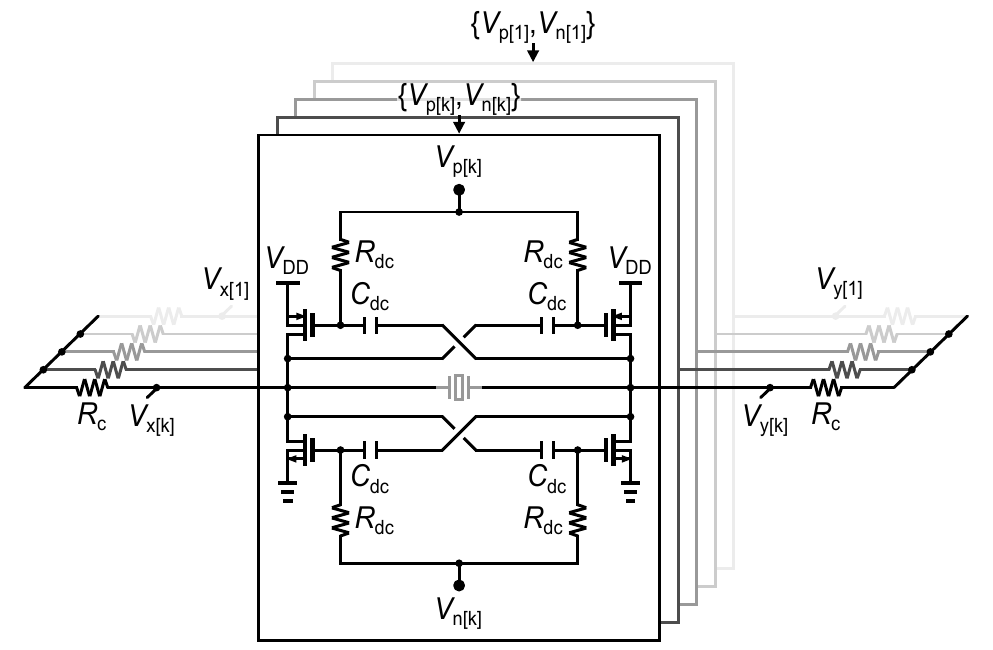}
\captionof{figure}{{\color{myred}\textbf{Overall circuit.}} CMOS implementation for the coupled oscillators based on the cross-coupled pairs.}\label{fig. S1}
\end{figure}

From the above expression, $V_{t}/I_{t}$ of the coupled oscillators consists of three active elements connected in series{\textemdash}a negative resistance $R_{tc}$, a negative inductance $L_{nc}$, and a negative capacitance $C_{nc}$. Each element is given as follows:

\begin{align}
R_{tc}&=\frac{-2}{\sum_{\alpha=1}^{k}G_{m[\alpha]}}\\
L_{nc}&=\frac{-2C_{z}k}{\left(\sum_{\alpha=1}^{k}G_{m[\alpha]}\right)^2}\\
C_{nc}&=\frac{-\sum_{\alpha=1}^{k}G_{m[\alpha]}R_{dc}C_{dc}}{2}
\end{align} 
\noindent\textbf{\color{myblue}\textit{A}.3.\space{\,}CMOS Implementation for the Coupled Oscillators}\\
{\indent}The frequency synchronization circuit model is implemented by connecting the CMOS-based \textit{2-Port Oscillator}s in parallel, as depicted in Fig. \ref{fig. S1}. The overall circuit in Fig. \ref{fig. S1} corresponds to the coupled oscillators shown in Fig. \ref{fig. 4} (without the parallel RLC model) or Fig. \ref{fig.11} (with the parallel RLC model). The circuit of the oscillation unit in Fig. \ref{fig. S1} is the same as the circuit depicted in Fig. \ref{fig. 2}(a), and its small-signal model is drawn in Fig. \ref{fig. 3} (without the parallel RLC model) or Fig. \ref{fig.8} (with the parallel RLC model). The two $p$-type MOSFETs in Fig. \ref{fig. S1} represent the two $g_{mp}$ of the upper cross-coupled pair in Fig. \ref{fig. 2}(a), and the two $n$-type MOSFETs in Fig. \ref{fig. S1} represent the two $g_{mn}$ of the lower cross-coupled pair in Fig. \ref{fig. 2}(a). The parasitic capacitances at the drain terminals of the $p$- and $n$-type MOSFETs represent $C_{z}$ depicted in Fig. \ref{fig. 2}(a). Therefore, $C_{z}$ is affected by the device size of the MOSFET. A higher value of $R_{c}$ disturbs the coupling between the oscillation units, whereas a lower value of $R_{c}$ improves synchronization characteristics.

The analysis and verification of the frequency synchronization circuit model are conducted using Cadence Virtuoso Studio (IC23.1-64b.43). In the overall circuit above, the design parameters and device sizes are summarized as follows:\\
{$\bullet\,\,$}$130{\,}\mathrm{nm}$ CMOS technology.\\
{$\bullet\,\,$}$V_{DD}=1.2{\,}\mathrm{V}$.\\
{$\bullet\,\,$}$V_{p[1]}${\textendash}$V_{p[k]}$ change from $0{\,}\mathrm{V}$ to $1.2{\,}\mathrm{V}$ (a higher $g_{mp}$ is achieved at a lower potential).\\
{$\bullet\,\,$}$V_{n[1]}${\textendash}$V_{n[k]}$ change from $0{\,}\mathrm{V}$ to $1.2{\,}\mathrm{V}$ (a higher $g_{mn}$ is achieved at a higher potential).\\
{$\bullet\,\,$}$R_{dc}=1{\,}\mathrm{k}{\Omega}$ and $C_{dc}=500{\,}\mathrm{fF}$.\\
{$\bullet\,\,$}$p$-type MOSFET size $=60\times(1{\,}\mathrm{um}/130{\,}\mathrm{nm})$.\\
{$\bullet\,\,$}$n$-type MOSFET size $=26\times(1{\,}\mathrm{um}/130{\,}\mathrm{nm})$.\\

To set the individual frequencies $f_{1}$ through $f_{8}$ in Figs. \ref{fig. 5} and \ref{fig.11}(b), $\{V_{p[\alpha]},V_{n[\alpha]}\}_{\alpha=1}^{8}$ are applied to the circuit shown in Fig. \ref{fig. S1} as follows:\\
{$\bullet\,\,$}$V_{p[1]}=0{\,}\mathrm{V}$, $V_{p[2]}=0.05{\,}\mathrm{V}$, $V_{p[3]}=0.1{\,}\mathrm{V}$, $V_{p[4]}=0.15{\,}\mathrm{V}$, $V_{p[5]}=0.2{\,}\mathrm{V}$, $V_{p[6]}=0.25{\,}\mathrm{V}$, $V_{p[7]}=0.3{\,}\mathrm{V}$, $V_{p[8]}=0.35{\,}\mathrm{V}$.\\
{$\bullet\,\,$}$V_{n[1]}=1.2{\,}\mathrm{V}$, $V_{n[2]}=1.15{\,}\mathrm{V}$, $V_{n[3]}=1.1{\,}\mathrm{V}$, $V_{n[4]}=1.05{\,}\mathrm{V}$, $V_{n[5]}=1{\,}\mathrm{V}$, $V_{n[6]}=0.95{\,}\mathrm{V}$, $V_{n[7]}=0.9{\,}\mathrm{V}$, $V_{n[8]}=0.85{\,}\mathrm{V}$.\\

{\noindent}Using the device sizes and bias conditions mentioned above, the $p$-type transconductance ($g_{mp}$) is typically obtained as $13.29{\,}\mathrm{mS}$, and the $n$-type transconductance ($g_{mn}$) is typically obtained as $14.07{\,}\mathrm{mS}$.{\singlespacing} 
\noindent\textbf{\color{myblue}\textit{A}.4.\space{\,}Details of the Small-Signal Analysis for the \textit{2-Port Oscillator} Involving a Resonator Model}\\
{\indent}Using $V'_{xp}=V'_{xn}=sR_{dc}C_{dc}V_{x}/(1+sR_{dc}C_{dc})$, $V'_{yp}=V'_{yn}=sR_{dc}C_{dc}V_{y}/(1+sR_{dc}C_{dc})$, and $Z_{dc}=(1+sR_{dc}C_{dc})/sC_{dc}$, $I_{t}$ and $-I_{t}$ in Equations (26) and (27) are rewritten as

\begin{align}
I_{t}&=\underbrace{g_{mp}V'_{yp}+\frac{V_{x}}{Z_{dc}}}_{I_{xp}}+\underbrace{g_{mn}V'_{yn}+\frac{V_{x}}{Z_{dc}}}_{I_{xn}}+\underbrace{sC_{z}V_{x}}_{I_{xcz}}+\underbrace{\frac{2V_{x}}{R_{p}}+\frac{2V_{x}}{sL_{p}}+s2C_{p}V_{x}}_{I_{xRLC}}\\
&=\left(g_{mp}+g_{mn}\right)V_{y}\frac{sR_{dc}C_{dc}}{1+sR_{dc}C_{dc}}+\frac{sC_{dc}2V_{x}}{1+sR_{dc}C_{dc}}+sC_{z}V_{x}+\frac{2V_{x}}{R_{p}}+\frac{2V_{x}}{sL_{p}}+s2C_{p}V_{x}\\
-I_{t}&=\underbrace{g_{mp}V'_{xp}+\frac{V_{y}}{Z_{dc}}}_{I_{yp}}+\underbrace{g_{mn}V'_{xn}+\frac{V_{y}}{Z_{dc}}}_{I_{yn}}+\underbrace{sC_{z}V_{y}}_{I_{ycz}}+\underbrace{\frac{2V_{y}}{R_{p}}+\frac{2V_{y}}{sL_{p}}+s2C_{p}V_{y}}_{I_{yRLC}}\\
&=\left(g_{mp}+g_{mn}\right)V_{x}\frac{sR_{dc}C_{dc}}{1+sR_{dc}C_{dc}}+\frac{sC_{dc}2V_{y}}{1+sR_{dc}C_{dc}}+sC_{z}V_{y}+\frac{2V_{y}}{R_{p}}+\frac{2V_{y}}{sL_{p}}+s2C_{p}V_{y}
\end{align}

{\noindent}Subtracting the above currents $I_{t}$ and $-I_{t}$, the result is given by

\begin{align}
2I_{t}=\left(g_{mp}+g_{mn}\right)\left(V_{y}-V_{x}\right)\frac{sR_{dc}C_{dc}}{1+sR_{dc}C_{dc}}+\frac{sC_{dc}2\left(V_{x}-V_{y}\right)}{1+sR_{dc}C_{dc}}+\left(sC_{z}+\frac{2}{R_{p}}+\frac{2}{sL_{p}}+s2C_{p}\right)\left(V_{x}-V_{y}\right)
\end{align}

{\noindent}Using $V_{t}=V_{x}-V_{y}$, the above expression is rewritten as

\begin{align}
2I_{t}=-\left(g_{mp}+g_{mn}\right)V_{t}\frac{sR_{dc}C_{dc}}{1+sR_{dc}C_{dc}}+\frac{sC_{dc}2V_{t}}{1+sR_{dc}C_{dc}}+\left(sC_{z}+\frac{2}{R_{p}}+\frac{2}{sL_{p}}+s2C_{p}\right)V_{t}
\end{align}

Multiplying the both sides of the above equation by $sR_{p}L_{p}\left(1+sR_{dc}C_{dc}\right)$, then rearranging that result for $V_{t}/I_{t}$, the impedance between nodes $X$ and $Y$ shown in Fig. \ref{fig.8} is obtained as follows:

\begin{align}
s2R_{p}L_{p}\left(1+sR_{dc}C_{dc}\right)I_{t}=&-\left(g_{mp}+g_{mn}\right)s^2R_{dc}C_{dc}R_{p}L_{p}V_{t}+{s^2}2C_{dc}R_{p}L_{p}V_{t}\\
{\nonumber}&+\left(s^2C_{z}R_{p}L_{p}+s2L_{p}+2R_{p}+{s^2}2R_{p}L_{p}C _{p}\right)\left(1+sR_{dc}C_{dc}\right)V_{t}
\end{align}

\begin{align}
\frac{V_{t}}{I_{t}}=\frac{s2R_{p}L_{p}\left(1+sR_{dc}C_{dc}\right)}{\left[\begin{array}{c}
-\left(g_{mp}+g_{mn}\right)s^2R_{dc}C_{dc}R_{p}L_{p}+{s^2}2C_{dc}R_{p}L_{p}+s^2R_{p}L_{p}\left(C_{z}+2C_{p}\right)\left(1+sR_{dc}C_{dc}\right)\\
+2\left(R_{p}+sL_{p}\right)\left(1+sR_{dc}C_{dc}\right)
\end{array}\right]}
\end{align}

{\noindent}Using $s=j{\omega}$, the above expression is rewritten as

\begin{align}
\frac{V_{t}}{I_{t}}=\frac{2R_{p}L_{p}\left(-{\omega}^2R_{dc}C_{dc}+j{\omega}\right)}{\left\{\begin{array}{c}
\left(g_{mp}+g_{mn}\right){\omega}^2R_{dc}C_{dc}R_{p}L_{p}-{\omega}^2{2}C_{dc}R_{p}L_{p}-{\omega}^2R_{p}L_{p}\left(C_{z}+2C_{p}\right)-{\omega}^2{2}R_{dc}C_{dc}L_{p}+2R_{p}\\
+j\left[-{\omega}^3R_{dc}C_{dc}R_{p}L_{p}\left(C_{z}+2C_{p}\right)+{\omega}2R_{dc}C_{dc}R_{p}+{\omega}2L_{p}\right]
\end{array}\right\}}
\end{align}

{\noindent}Given that the order of the design parameters used in Fig. \ref{fig. S1}, the real part of the denominator in the above expression can be approximated as follows:

\begin{align}
\frac{V_{t}}{I_{t}}{\,\approx\,}\frac{2R_{p}L_{p}\left(-{\omega}^2R_{dc}C_{dc}+j{\omega}\right)}{\left(g_{mp}+g_{mn}\right){\omega}^2R_{dc}C_{dc}R_{p}L_{p}+j\left[-{\omega}^3R_{dc}C_{dc}R_{p}L_{p}\left(C_{z}+2C_{p}\right)+{\omega}2R_{dc}C_{dc}R_{p}+{\omega}2L_{p}\right]}
\end{align}

Before solving the above equation, let us recall that the resonance of a series (or parallel) RLC circuit occurs when the reactance of its impedance becomes zero \cite{Lee2026osc}. Therefore, the resonant frequency from the above equation can be derived by finding the reactance $X'_{n}$ of $V_{t}/I_{t}$ and solving $X'_{n}=0$. To find $X'_{n}$, the numerator and denominator of $V_{t}/I_{t}$ are multiplied by the complex conjugate of the denominator as follows:

\begin{align}
\begin{gathered}
\frac{2R_{p}L_{p}\left(-{\omega}^2R_{dc}C_{dc}+j{\omega}\right)\left\{\left(g_{mp}+g_{mn}\right){\omega}^2R_{dc}C_{dc}R_{p}L_{p}+j\left[{\omega}^3R_{dc}C_{dc}R_{p}L_{p}\left(C_{z}+2C_{p}\right)-{\omega}2R_{dc}C_{dc}R_{p}-{\omega}2L_{p}\right]\right\}}{\left[\left(g_{mp}+g_{mn}\right){\omega}^2R_{dc}C_{dc}R_{p}L_{p}\right]^2+\left[-{\omega}^3R_{dc}C_{dc}R_{p}L_{p}\left(C_{z}+2C_{p}\right)+{\omega}2R_{dc}C_{dc}R_{p}+{\omega}2L_{p}\right]^2}\\
{\big\downarrow}\\
R'_{t}+jX'_{n}
\end{gathered}
\end{align}

{\noindent}Then, from the above expression, the real part (or resistance $R'_{t}$) and imaginary part (or reactance $X'_{n}$) of $V_{t}/I_{t}$ are obtained as follows:

\begin{align}
R'_{t}&=\frac{2R_{p}L_{p}\left[-\left(g_{mp}+g_{mn}\right){\omega}^4R_{dc}^2C_{dc}^2R_{p}L_{p}-{\omega}^4R_{dc}C_{dc}R_{p}L_{p}\left(C_{z}+2C_{p}\right)+{\omega}^2{2}R_{dc}C_{dc}R_{p}+{\omega}^2{2}L_{p}\right]}{\left[\left(g_{mp}+g_{mn}\right){\omega}^2R_{dc}C_{dc}R_{p}L_{p}\right]^2+\left[-{\omega}^3R_{dc}C_{dc}R_{p}L_{p}\left(C_{z}+2C_{p}\right)+{\omega}2R_{dc}C_{dc}R_{p}+{\omega}2L_{p}\right]^2}\\
X'_{n}&=\frac{2R_{p}L_{p}\left[\left(g_{mp}+g_{mn}\right){\omega}^3R_{dc}C_{dc}R_{p}L_{p}-{\omega}^5R_{dc}^2C_{dc}^2R_{p}L_{p}\left(C_{z}+2C_{p}\right)+{\omega}^3{2}R_{dc}^2C_{dc}^2R_{p}+{\omega}^3{2}R_{dc}C_{dc}L_{p}\right]}{\left[\left(g_{mp}+g_{mn}\right){\omega}^2R_{dc}C_{dc}R_{p}L_{p}\right]^2+\left[-{\omega}^3R_{dc}C_{dc}R_{p}L_{p}\left(C_{z}+2C_{p}\right)+{\omega}2R_{dc}C_{dc}R_{p}+{\omega}2L_{p}\right]^2}
\end{align}

{\noindent}For solving $X'_{n}=0$, the numerator of $X'_{n}$ can be written as

\begin{align}
\left(g_{mp}+g_{mn}\right){\omega}^3R_{dc}C_{dc}R_{p}L_{p}-{\omega}^5R_{dc}^2C_{dc}^2R_{p}L_{p}\left(C_{z}+2C_{p}\right)+{\omega}^3{2}R_{dc}^2C_{dc}^2R_{p}+{\omega}^3{2}R_{dc}C_{dc}L_{p}=0
\end{align}

{\noindent}Then, {$\omega$} (or $2{\pi}f$) is obtained as follows:

\begin{align}
{\omega}^5R_{dc}^2C_{dc}^2R_{p}L_{p}\left(C_{z}+2C_{p}\right)&=\left(g_{mp}+g_{mn}\right){\omega}^3R_{dc}C_{dc}R_{p}L_{p}+{\omega}^3{2}R_{dc}^2C_{dc}^2R_{p}+{\omega}^3{2}R_{dc}C_{dc}L_{p}\\
{\omega}^2&=\frac{\left(g_{mp}+g_{mn}\right)R_{dc}C_{dc}R_{p}L_{p}+2R_{dc}^2C_{dc}^2R_{p}+2R_{dc}C_{dc}L_{p}}{R_{dc}^2C_{dc}^2R_{p}L_{p}\left(C_{z}+2C_{p}\right)}\\
{\omega}&=\sqrt{\frac{g_{mp}+g_{mn}}{R_{dc}C_{dc}\left(C_{z}+2C_{p}\right)}+\frac{2}{L_{p}\left(C_{z}+2C_{p}\right)}+\frac{2}{R_{dc}C_{dc}R_{p}\left(C_{z}+2C_{p}\right)}}\\
&=\sqrt{\frac{G_{m}}{R_{dc}C_{dc}C_{xy}}+\frac{2}{L_{p}C_{xy}}+\frac{2}{R_{dc}C_{dc}R_{p}C_{xy}}}
\end{align}

{\noindent}where $G_{m}=g_{mp}+g_{mn}$ and $C_{xy}=C_{z}+2C_{p}$. Therefore, the resonant frequency of the \textit{2-Port Oscillator} involving a resonator model is derived as follows:

\begin{align}
f&=\frac{1}{2{\pi}}\sqrt{\frac{G_{m}}{R_{dc}C_{dc}C_{xy}}+\frac{2}{L_{p}C_{xy}}+\frac{2}{R_{dc}C_{dc}R_{p}C_{xy}}}\\
&=\frac{1}{2{\pi}}\sqrt{\frac{1}{R_{dc}C_{dc}R_{Gm}C_{xy}}+\frac{2}{L_{p}C_{xy}}+\frac{2}{R_{dc}C_{dc}R_{p}C_{xy}}}
\end{align}

{\noindent}where $G_{m}=1/R_{Gm}$.{\singlespacing} 
\noindent\textbf{\color{myblue}\textit{A}.5.\space{\,}Details of the Small-Signal Analysis for the Coupled Oscillators Involving Resonator Models}\\
{\indent}Using $V'_{xp}=V'_{xn}=sR_{dc}C_{dc}V_{x}/(1+sR_{dc}C_{dc})$, $V'_{yp}=V'_{yn}=sR_{dc}C_{dc}V_{y}/(1+sR_{dc}C_{dc})$, and $Z_{dc}=(1+sR_{dc}C_{dc})/sC_{dc}$, $I_{t}$ and $-I_{t}$ in Equations (35) and (36) are rewritten as

\begin{align}
I_{t}&=\underbrace{\sum_{\alpha=1}^{k}g_{mp[\alpha]}V'_{yp}+\frac{kV_{x}}{Z_{dc}}}_{k{\cdot}I_{xp}}+\underbrace{\sum_{\alpha=1}^{k}g_{mn[\alpha]}V'_{yn}+\frac{kV_{x}}{Z_{dc}}}_{k{\cdot}I_{xn}}+\underbrace{sC_{z}kV_{x}}_{k{\cdot}I_{xcz}}+\underbrace{\frac{2kV_{x}}{R_{p}}+\frac{2kV_{x}}{sL_{p}}+s2C_{p}kV_{x}}_{k{\cdot}I_{xRLC}}\\
{\nonumber}&=\sum_{\alpha=1}^{k}G_{m[\alpha]}V_{y}\frac{sR_{dc}C_{dc}}{1+sR_{dc}C_{dc}}+\frac{sC_{dc}2kV_{x}}{1+sR_{dc}C_{dc}}+\left(sC_{z}+\frac{2}{R_{p}}+\frac{2}{sL_{p}}+s2C_{p}\right)kV_{x}
\end{align}

\begin{align}
-I_{t}&=\underbrace{\sum_{\alpha=1}^{k}g_{mp[\alpha]}V'_{xp}+\frac{kV_{y}}{Z_{dc}}}_{k{\cdot}I_{yp}}+\underbrace{\sum_{\alpha=1}^{k}g_{mn[\alpha]}V'_{xn}+\frac{kV_{y}}{Z_{dc}}}_{k{\cdot}I_{yn}}+\underbrace{sC_{z}kV_{y}}_{k{\cdot}I_{ycz}}+\underbrace{\frac{2kV_{y}}{R_{p}}+\frac{2kV_{y}}{sL_{p}}+s2C_{p}kV_{y}}_{k{\cdot}I_{yRLC}}\\
{\nonumber}&=\sum_{\alpha=1}^{k}G_{m[\alpha]}V_{x}\frac{sR_{dc}C_{dc}}{1+sR_{dc}C_{dc}}+\frac{sC_{dc}2kV_{y}}{1+sR_{dc}C_{dc}}+\left(sC_{z}+\frac{2}{R_{p}}+\frac{2}{sL_{p}}+s2C_{p}\right)kV_{y}
\end{align}

{\noindent}where $G_{m[\alpha]}=g_{mp[\alpha]}+g_{mn[\alpha]}$, and $k$ is the number of coupled oscillators. Then, the subtraction of $I_{t}$ and $-I_{t}$ is

\begin{align}
2I_{t}=\sum_{\alpha=1}^{k}G_{m[\alpha]}\frac{sR_{dc}C_{dc}}{1+sR_{dc}C_{dc}}\left(V_{y}-V_{x}\right)+\frac{sC_{dc}2k\left(V_{x}-V_{y}\right)}{1+sR_{dc}C_{dc}}+\left(sC_{z}+\frac{2}{R_{p}}+\frac{2}{sL_{p}}+s2C_{p}\right)k\left(V_{x}-V_{y}\right)
\end{align}

{\noindent}Applying $V_{t}=V_{x}-V_{y}$ to the above equation and rearranging that result for $V_{t}/I_{t}$, the impedance is expressed as follows:

\begin{align}
2I_{t}&=-\sum_{\alpha=1}^{k}G_{m[\alpha]}\frac{sR_{dc}C_{dc}}{1+sR_{dc}C_{dc}}V_{t}+\frac{sC_{dc}2kV_{t}}{1+sR_{dc}C_{dc}}+\left(sC_{z}+\frac{2}{R_{p}}+\frac{2}{sL_{p}}+s2C_{p}\right)kV_{t}\\
\frac{V_{t}}{I_{t}}&=\frac{s2R_{p}L_{p}\left(1+sR_{dc}C_{dc}\right)}{-\sum_{\alpha=1}^{k}G_{m[\alpha]}s^2R_{dc}C_{dc}R_{p}L_{p}+s^2C_{dc}2kR_{p}L_{p}+\left(s^2C_{z}R_{p}L_{p}+s2L_{p}+2R_{p}+s^2{2}R_{p}L_{p}C_{p}\right)k\left(1+sR_{dc}C_{dc}\right)}\\
&=\frac{s2R_{p}L_{p}\left(1+sR_{dc}C_{dc}\right)}{-\sum_{\alpha=1}^{k}G_{m[\alpha]}s^2R_{dc}C_{dc}R_{p}L_{p}+s^2C_{dc}2kR_{p}L_{p}+\left[s^2R_{p}L_{p}\left(C_{z}+2C_{p}\right)+2\left(R_{p}+sL_{p}\right)\right]k\left(1+sR_{dc}C_{dc}\right)}\\
&=\frac{s2R_{p}L_{p}\left(1+sR_{dc}C_{dc}\right)}{\left[\begin{array}{c}
-\sum_{\alpha=1}^{k}G_{m[\alpha]}s^2R_{dc}C_{dc}R_{p}L_{p}+s^2C_{dc}2kR_{p}L_{p}+s^2R_{p}L_{p}\left(C_{z}+2C_{p}\right)k\left(1+sR_{dc}C_{dc}\right)\\
+2k\left(R_{p}+sL_{p}\right)\left(1+sR_{dc}C_{dc}\right)
\end{array}\right]}
\end{align}

{\noindent}Using $s=j{\omega}$, the above expression is rewritten as

\begin{align}
\frac{V_{t}}{I_{t}}&=\frac{2R_{p}L_{p}\left(-{\omega}^2R_{dc}C_{dc}+j{\omega}\right)}{\left[\begin{array}{c}
\sum_{\alpha=1}^{k}G_{m[\alpha]}{\omega}^2R_{dc}C_{dc}R_{p}L_{p}-{\omega}^2C_{dc}2kR_{p}L_{p}-{\omega}^2R_{p}L_{p}\left(C_{z}+2C_{p}\right)k-j{\omega}^3R_{dc}C_{dc}R_{p}L_{p}\left(C_{z}+2C_{p}\right)k\\
2kR_{p}-2k{\omega}^2R_{dc}C_{dc}L_{p}+j{\omega}2kL_{p}+j{\omega}2kR_{dc}C_{dc}R_{p}
\end{array}\right]}\\
&=\frac{2R_{p}L_{p}\left(-{\omega}^2R_{dc}C_{dc}+j{\omega}\right)}{\left\{\begin{array}{c}
\sum_{\alpha=1}^{k}G_{m[\alpha]}{\omega}^2R_{dc}C_{dc}R_{p}L_{p}-{\omega}^2C_{dc}2kR_{p}L_{p}-{\omega}^2R_{p}L_{p}\left(C_{z}+2C_{p}\right)k-2k{\omega}^2R_{dc}C_{dc}L_{p}+2kR_{p}\\
+j\left[-{\omega}^3R_{dc}C_{dc}R_{p}L_{p}\left(C_{z}+2C_{p}\right)k+{\omega}2kR_{dc}C_{dc}R_{p}+{\omega}2kL_{p}\right]
\end{array}\right\}}
\end{align}

As with the approximation of $V_{t}/I_{t}$ performed in Appendix A.4, given that the order of the design parameters used in Fig. \ref{fig. S1}, the real part of the above denominator can be approximated as follows:

\begin{align}
\frac{V_{t}}{I_{t}}&{\,\approx\,}\frac{2R_{p}L_{p}\left(-{\omega}^2R_{dc}C_{dc}+j{\omega}\right)}{\sum_{\alpha=1}^{k}G_{m[\alpha]}{\omega}^2R_{dc}C_{dc}R_{p}L_{p}+j\left[-{\omega}^3R_{dc}C_{dc}R_{p}L_{p}\left(C_{z}+2C_{p}\right)k+{\omega}2kR_{dc}C_{dc}R_{p}+{\omega}2kL_{p}\right]}\\
&=\frac{2R_{p}L_{p}\left(-{\omega}^2R_{dc}C_{dc}+j{\omega}\right)\left\{\sum_{\alpha=1}^{k}G_{m[\alpha]}{\omega}^2R_{dc}C_{dc}R_{p}L_{p}+j\left[{\omega}^3R_{dc}C_{dc}R_{p}L_{p}\left(C_{z}+2C_{p}\right)-{\omega}2R_{dc}C_{dc}R_{p}-{\omega}2L_{p}\right]k\right\}}{\left(\sum_{\alpha=1}^{k}G_{m[\alpha]}{\omega}^2R_{dc}C_{dc}R_{p}L_{p}\right)^2+\left[-{\omega}^3R_{dc}C_{dc}R_{p}L_{p}\left(C_{z}+2C_{p}\right)+{\omega}2R_{dc}C_{dc}R_{p}+{\omega}2L_{p}\right]^2k^2}\\
&=R'_{tc}+jX'_{nc}
\end{align}

{\noindent}where $R'_{tc}$ and $X'_{nc}$ are the real part (or resistance) and the imaginary part (or reactance) of $V_{t}/I_{t}$, respectively. The complete expression for $R'_{tc}$ and $X'_{nc}$ are given by

\begin{align}
R'_{tc}&=\frac{2R_{p}L_{p}\left\{-\sum_{\alpha=1}^{k}G_{m[\alpha]}{\omega}^4R_{dc}^2C_{dc}^2R_{p}L_{p}-\left[{\omega}^4R_{dc}C_{dc}R_{p}L_{p}\left(C_{z}+2C_{p}\right)-{\omega}^2{2}R_{dc}C_{dc}R_{p}-{\omega}^2{2}L_{p}\right]k\right\}}{\left(\sum_{\alpha=1}^{k}G_{m[\alpha]}{\omega}^2R_{dc}C_{dc}R_{p}L_{p}\right)^2+\left[-{\omega}^3R_{dc}C_{dc}R_{p}L_{p}\left(C_{z}+2C_{p}\right)+{\omega}2R_{dc}C_{dc}R_{p}+{\omega}2L_{p}\right]^2k^2}\\
X'_{nc}&=\frac{2R_{p}L_{p}\left\{\sum_{\alpha=1}^{k}G_{m[\alpha]}{\omega}^3R_{dc}C_{dc}R_{p}L_{p}-\left[{\omega}^5R_{dc}^2C_{dc}^2R_{p}L_{p}\left(C_{z}+2C_{p}\right)-{\omega}^3{2}R_{dc}^2C_{dc}^2R_{p}-{\omega}^3{2}R_{dc}C_{dc}L_{p}\right]k\right\}}{\left(\sum_{\alpha=1}^{k}G_{m[\alpha]}{\omega}^2R_{dc}C_{dc}R_{p}L_{p}\right)^2+\left[-{\omega}^3R_{dc}C_{dc}R_{p}L_{p}\left(C_{z}+2C_{p}\right)+{\omega}2R_{dc}C_{dc}R_{p}+{\omega}2L_{p}\right]^2k^2}
\end{align}

As conducted in Appendix A.4 to derive a resonant frequency from $R'_{t}+jX'_{n}$, the resonant frequency of the coupled oscillators involving resonator models can be derived by solving $X'_{nc}=0$. Note that a series (or parallel) RLC circuit has its resonant peak when the reactance (or imaginary part) of its impedance becomes zero \cite{Lee2026osc}. Therefore, the resonant frequency can be derived from $R'_{tc}+jX'_{nc}$ by solving $X'_{nc}=0$ as follows:

\begin{align}
\sum_{\alpha=1}^{k}G_{m[\alpha]}{\omega}^3R_{dc}C_{dc}R_{p}L_{p}-\left[{\omega}^5R_{dc}^2C_{dc}^2R_{p}L_{p}\left(C_{z}+2C_{p}\right)-{\omega}^3{2}R_{dc}^2C_{dc}^2R_{p}-{\omega}^3{2}R_{dc}C_{dc}L_{p}\right]k=0
\end{align}

{\noindent}Then, ${\omega}$ (or $2{\pi}f$) is obtained as follows:

\begin{align}
{\omega}^5R_{dc}^2C_{dc}^2R_{p}L_{p}\left(C_{z}+2C_{p}\right)k&=\sum_{\alpha=1}^{k}G_{m[\alpha]}{\omega}^3R_{dc}C_{dc}R_{p}L_{p}+{\omega}^3{2}R_{dc}^2C_{dc}^2R_{p}k+{\omega}^3{2}R_{dc}C_{dc}L_{p}k\\
{\omega}^2&=\frac{\sum_{\alpha=1}^{k}G_{m[\alpha]}R_{dc}C_{dc}R_{p}L_{p}+2R_{dc}^2C_{dc}^2R_{p}k+2R_{dc}C_{dc}L_{p}k}{R_{dc}^2C_{dc}^2R_{p}L_{p}\left(C_{z}+2C_{p}\right)k}\\
{\omega}&=\sqrt{\frac{\sum_{\alpha=1}^{k}G_{m[\alpha]}}{R_{dc}C_{dc}\left(C_{z}+2C_{p}\right)k}+\frac{2}{L_{p}\left(C_{z}+2C_{p}\right)}+\frac{2}{R_{dc}C_{dc}R_{p}\left(C_{z}+2C_{p}\right)}}
\end{align}

{\noindent}The resonant frequency is

\begin{align}
f_{syn}&=\frac{1}{2{\pi}}\sqrt{\frac{\sum_{\alpha=1}^{k}G_{m[\alpha]}}{R_{dc}C_{dc}\left(C_{z}+2C_{p}\right)k}+\frac{2}{L_{p}\left(C_{z}+2C_{p}\right)}+\frac{2}{R_{dc}C_{dc}R_{p}\left(C_{z}+2C_{p}\right)}}\\
&=\frac{1}{2{\pi}}\sqrt{\frac{\sum_{\alpha=1}^{k}G_{m[\alpha]}}{R_{dc}C_{dc}C_{xy}k}+\frac{2}{L_{p}C_{xy}}+\frac{2}{R_{dc}C_{dc}R_{p}C_{xy}}}\\
&=\frac{1}{2{\pi}}\sqrt{\sum_{\alpha=1}^{k}\frac{1}{R_{Gm[\alpha]}C_{xy}kR_{dc}C_{dc}}+\frac{2}{L_{p}C_{xy}}+\frac{2}{R_{dc}C_{dc}R_{p}C_{xy}}}
\end{align}

{\noindent}where $C_{xy}=C_{z}+2C_{p}$ and $G_{m[\alpha]}=1/R_{Gm[\alpha]}$.

{\singlespacing}If an oscillator in the coupled model shown in Fig. \ref{fig.10} has an individually controllable $C_{z}$ and $\{R_{p[\alpha]},L_{p[\alpha]},C_{p[\alpha]}\}_{\alpha=1}^{k}$, the two small-signal currents in Equations (35) and (36) can be modified as follows:

\begin{align}
I_{t}&=\underbrace{\sum_{\alpha=1}^{k}g_{mp[\alpha]}V'_{yp}+\frac{kV_{x}}{Z_{dc}}}_{k{\cdot}I_{xp}}+\underbrace{\sum_{\alpha=1}^{k}g_{mn[\alpha]}V'_{yn}+\frac{kV_{x}}{Z_{dc}}}_{k{\cdot}I_{xn}}+\underbrace{sC'_{z}V_{x}}_{I'_{xcz}}+\underbrace{\frac{2V_{x}}{R'_{p}}+\frac{2V_{x}}{sL'_{p}}+s2C'_{p}V_{x}}_{I'_{xRLC}}\\
-I_{t}&=\underbrace{\sum_{\alpha=1}^{k}g_{mp[\alpha]}V'_{xp}+\frac{kV_{y}}{Z_{dc}}}_{k{\cdot}I_{yp}}+\underbrace{\sum_{\alpha=1}^{k}g_{mn[\alpha]}V'_{xn}+\frac{kV_{y}}{Z_{dc}}}_{k{\cdot}I_{yn}}+\underbrace{sC'_{z}V_{y}}_{I'_{ycz}}+\underbrace{\frac{2V_{y}}{R'_{p}}+\frac{2V_{y}}{sL'_{p}}+s2C'_{p}V_{y}}_{I'_{yRLC}}
\end{align}

{\noindent}Compared to Equations (35) and (36), the model parameters are updated as follows:

\begin{align}
kC_{z}&{\,\rightarrow\,}C'_{z}=\sum_{\alpha=1}^{k}C_{z[\alpha]}\\
\frac{R_{p}}{k}&{\,\rightarrow\,}R'_{p}=R_{p[1]}||R_{p[2]}{\cdots}||R_{p[k-1]}||R_{p[k]}\\
\frac{L_{p}}{k}&{\,\rightarrow\,}L'_{p}=L_{p[1]}||L_{p[2]}{\cdots}||L_{p[k-1]}||L_{p[k]}\\
kC_{p}&{\,\rightarrow\,}C'_{p}=\sum_{\alpha=1}^{k}C_{p[\alpha]}
\end{align}

{\noindent}where $k$ is the number of coupled oscillators. Then, subtracting the modified two currents above and applying $V_{t}=V_{x}-V_{y}$ and $G_{m[\alpha]}=g_{mp[\alpha]}+g_{mn[\alpha]}$ to that result, the following equation is given by

\begin{align}
2I_{t}=-\sum_{\alpha=1}^{k}G_{m[\alpha]}\frac{sR_{dc}C_{dc}}{1+sR_{dc}C_{dc}}V_{t}+\frac{sC_{dc}2kV_{t}}{1+sR_{dc}C_{dc}}+\left(sC'_{z}+\frac{2}{R'_{p}}+\frac{2}{sL'_{p}}+s2C'_{p}\right)V_{t}
\end{align}

{\noindent}Rearranging the above equation for $V_{t}/I_{t}$, the following equations are given by

\begin{align}
\frac{V_{t}}{I_{t}}&=\frac{s2R'_{p}L'_{p}\left(1+sR_{dc}C_{dc}\right)}{-\sum_{\alpha=1}^{k}G_{m[\alpha]}s^2R_{dc}C_{dc}R'_{p}L'_{p}+s^2C_{dc}2kR'_{p}L'_{p}+\left(s^2C'_{z}R'_{p}L'_{p}+s2L'_{p}+2R'_{p}+s^2{2}R'_{p}L'_{p}C'_{p}\right)\left(1+sR_{dc}C_{dc}\right)}\\
&=\frac{s2R'_{p}L'_{p}\left(1+sR_{dc}C_{dc}\right)}{\left[\begin{array}{c}
-\sum_{\alpha=1}^{k}G_{m[\alpha]}s^2R_{dc}C_{dc}R'_{p}L'_{p}+s^2C_{dc}2kR'_{p}L'_{p}+s^2R'_{p}L'_{p}\left(C'_{z}+2C'_{p}\right)\left(1+sR_{dc}C_{dc}\right)\\
+2\left(R'_{p}+sL'_{p}\right)\left(1+sR_{dc}C_{dc}\right)
\end{array}\right]}\\
&=\frac{2R'_{p}L'_{p}\left(-{\omega}^2R_{dc}C_{dc}+j{\omega}\right)}{\left\{\begin{array}{c}
\sum_{\alpha=1}^{k}G_{m[\alpha]}{\omega}^2R_{dc}C_{dc}R'_{p}L'_{p}-{\omega}^2C_{dc}2kR'_{p}L'_{p}-{\omega}^2R'_{p}L'_{p}\left(C'_{z}+2C'_{p}\right)-2{\omega}^2R_{dc}C_{dc}L'_{p}+2R'_{p}\\
+j\left[-{\omega}^3R_{dc}C_{dc}R'_{p}L'_{p}\left(C'_{z}+2C'_{p}\right)+{\omega}2R_{dc}C_{dc}R'_{p}+{\omega}2L'_{p}\right]
\end{array}\right\}}\\
&{\,\approx\,}\frac{2R'_{p}L'_{p}\left(-{\omega}^2R_{dc}C_{dc}+j{\omega}\right)}{\sum_{\alpha=1}^{k}G_{m[\alpha]}{\omega}^2R_{dc}C_{dc}R'_{p}L'_{p}+j\left[-{\omega}^3R_{dc}C_{dc}R'_{p}L'_{p}\left(C'_{z}+2C'_{p}\right)+{\omega}2R_{dc}C_{dc}R'_{p}+{\omega}2L'_{p}\right]}\\
&=R''_{tc}+jX''_{nc}
\end{align}

Note that a series (or parallel) RLC circuit has its resonant peak when the reactance (or imaginary part) of its impedance becomes zero \cite{Lee2026osc}. By solving $X''_{nc}=0$ using the numerator of $X''_{nc}$, the resonant frequency can be derived as follows:

\begin{align}
\underbrace{\sum_{\alpha=1}^{k}G_{m[\alpha]}{\omega}^3R_{dc}C_{dc}R'_{p}L'_{p}-{\omega}^5R_{dc}^2C_{dc}^2R'_{p}L'_{p}\left(C'_{z}+2C'_{p}\right)+{\omega}^3{2}R_{dc}^2C_{dc}^2R'_{p}+{\omega}^3{2}R_{dc}C_{dc}L'_{p}}_{\text{Numerator of}{\,}X''_{nc}{\,}\text{for}X''_{nc}=0}=0
\end{align}

\begin{align}
f_{syn}&=\frac{1}{2{\pi}}\sqrt{\frac{\sum_{\alpha=1}^{k}G_{m[\alpha]}}{R_{dc}C_{dc}\left(C'_{z}+2C'_{p}\right)}+\frac{2}{L'_{p}\left(C'_{z}+2C'_{p}\right)}+\frac{2}{R_{dc}C_{dc}R'_{p}\left(C'_{z}+2C'_{p}\right)}}\\
&=\frac{1}{2{\pi}}\sqrt{\frac{\sum_{\alpha=1}^{k}G_{m[\alpha]}}{R_{dc}C_{dc}C'_{xy}}+\frac{2}{L'_{p}C'_{xy}}+\frac{2}{R_{dc}C_{dc}R'_{p}C'_{xy}}}\\
&=\frac{1}{2{\pi}}\sqrt{\sum_{\alpha=1}^{k}\frac{1}{R_{Gm[\alpha]}C'_{xy}R_{dc}C_{dc}}+\frac{2}{L'_{p}C'_{xy}}+\frac{2}{R_{dc}C_{dc}R'_{p}C'_{xy}}}
\end{align}

{\noindent}where $C'_{xy}=C'_{z}+2C'_{p}$ and $G_{m[\alpha]}=1/R_{Gm[\alpha]}$. 
\end{document}